\documentstyle[11pt,aaspp4,flushrt]{article}
\input epsf
\begin{document}
\def\gtsima{$\, \buildrel > \over \sim \,$}
\def\ltsima{$\, \buildrel < \over \sim \,$}
\def\simgt{\lower.5ex\hbox{\gtsima}}
\def\simlt{\lower.5ex\hbox{\ltsima}}
\def\sm{$\sim\,$}
\def\onesigma{$1\,\sigma$}
\def\zhel{\ifmmode z_\odot\else$z_\odot$\fi}
\def\nhat{\ifmmode {\hat{\bf n}}\else${\hat {\bf n}}$\fi}
\def\degs{\ifmmode^\circ\else$^\circ$\fi}
\def\degspt{\ifmmode^\circ_\cdot\else$^\circ_\cdot$\fi}
\def\mins{$'$\ }
\def\minspt{$'_\cdot$}
\def\secs{$''$}
\def\secspt{$''_\cdot$}
\def\hr{$^h$}
\def\hpt{$^h_\cdot$}
\def\m{$^m$\ }
\def\mpt{$^m_\cdot$}
\def\s{$^s$\ }
\def\spt{$^s_\cdot$}
\def\kms{\ifmmode{\rm km}\,{\rm s}^{-1}\else km$\,$s$^{-1}$\fi}
\def\ksmpc{\ifmmode{\rm km}\,{\rm s}^{-1}\,{\rm Mpc}^{-1}\else km$\,$s$^{-1}\,$Mpc$^{-1}$\fi}
\def\etal{{\sl et al.}}
\def\ie{{\it i.e.}}
\def\eg{{\it e.g.}}
\def\apriori{{\em a priori}}
\def\aposteriori{{\em a posteriori}}
\def\halpha{H$\alpha$}
\def\h1{$h^{-1}$}
\def\dnsigma{$D_n$-$\sigma$}
\font\tensm=cmcsc10
\def\hii{H\kern 2.0pt{\tensm ii}}	
\def\hi{\ifmmode{\rm H\kern 2.0pt{\tensm I}}\else H\kern 2.0pt{\tensm I}\fi}	
\def\iras{{\sl IRAS}}
\def\potent{{\small POTENT}}
\def\calr{{\cal R}}
\def\dinv{d_{{\rm inv}}}
\def\deltainv{\Delta_{{\rm inv}}}
\def\muinv{\mu(\dinv)}
\def\delinv{\Delta_{{\rm inv}}}
\def\onehalf{\frac{1}{2}}
\def\sigz{\sigma_z}
\def\sigeta{\sigma_{\eta}}
\def\sigr{\sigma_r}
\def\sigv{\sigma_v}
\def\sigin{\sigma_{{\rm in}}}
\def\sigout{\sigma_{{\rm out}}}
\def\sigM{\sigma_M}
\def\etazero{\eta^0}
\def\etapr{\eta^{0\prime}}
\def\nsigeta{\frac{1}{\sqrt{2\pi}\,\sigeta}}
\def\nsigr{\frac{1}{\sqrt{2\pi}\,\sigr}}
\def\nsigv{\frac{1}{\sqrt{2\pi}\,\sigv}}
\def\nsigM{\frac{1}{\sqrt{2\pi}\,\sigM}}
\def\nsig{\frac{1}{\sqrt{2\pi}\,\sigma}}
\def\nsigD{\frac{1}{\sqrt{2\pi}\,\Delta}}
\def\nsigDinv{\frac{1}{\sqrt{2\pi}\,\delinv}}
\def\nsigzero{{1 \over \sqrt{2\pi}\,\sigma_0}}
\def\nsigm0{{1 \over \sqrt{2\pi}\,\sigma_{0m}}}
\def\nsigxi{\frac{1}{\sqrt{2\pi}\,\sigma_\xi}}
\def\nsigz{\frac{1}{\sqrt{2\pi}\,\sigz}}
\def\2overpi{2 \over \pi}
\def\expmM{\exp\!\left(-\frac{(m-(M(\eta)+\mu(r)))^2}{2\sigma^2} \right)}
\def\expmMm0{\exp\!\left(-\frac{(m-(M(\eta_0)+\mu(r)))^2}{2\sigma_{0m}^2} \right)}
\def\expeta{e^{-\frac{(\eta-\eta_0)^2}{2\sigeta^2}}}
\def\expet{\exp\!\left(-\frac{[\eta - \eta^0(m-\mu(r))]^2}{2\sigeta^2}\right)}
\def\expetm{\exp\!\left(-\frac{[\eta - \eta^0(m-\mu)]^2}{2\sigeta^2}\right)}
\def\expmm{\exp\!\left(-\frac{(m-m(\eta))^2}{2\sigma^2}\right)}
\def\expmu{\exp\!\left(-\frac{(\mu(r) - \mu(m,\eta))^2}{2 \sigma^2} \right)}
\def\expmuD{\exp\!\left(-\frac{(\mu(r) - \mu(d))^2}{2 \sigma^2} \right)}
\def\explnrD{\exp\!\left(-\frac{\left[\ln r/d \right]^2}{2 \Delta^2} \right)}
\def\explnrDinv{\exp\!\left(-\frac{\left[\ln r/\dinv \right]^2}{2 \delinv^2} \right)}
\def\explnvD{e^{-\frac{\left[\ln\frac{v_c}{\vd}\right]^2}{2 \Delta^2}}}
\def\explnu{e^{-\frac{\left[\ln\frac{v_r-u}{v_c}\right]^2}{2 \Delta^2}}}
\def\explnx{e^{-\frac{\left(\ln x\right)^2}{2 \Delta^2}}}
\def\explnxinv{e^{-\frac{\left(\ln x\right)^2}{2 \delinv^2}}}
\def\expvr{\exp\!\left(-\frac{[v_r - (v_c+ v_p(v_c))]^2}{2\sigr^2}\right)}
\def\expcz{\exp\!\left(-\frac{[z - (r+ u(r))]^2}{2\sigv^2}\right)}
\def\expxi{\exp\!\left(-\frac{(\xi - \xi(m,\eta))^2}{2 \sigma_{\xi}^2} \right)}
\def\expz{\exp\!\left(-\frac{(m_z-m_z(m,\eta))^2}{2\sigz^2} \right)}
\def\xlim{\xi_\ell}
\def\xint{\int_{\xlim}^\infty\,}
\def\xxint{\int^{\xlim}_{-\infty}\,}
\def\erf{{\rm erf}}
\def\infint{\int_{-\infty}^\infty}
\def\inf0int{\int_0^\infty}
\def\mlim{m_\ell}
\def\mlint{\int_{-\infty}^{\mlim}\,}
\def\Aint{\int_{-\infty}^\cala\,}
\def\onplserf{\left[1+\erf\left(\frac{\xi(m,\eta)-\xlim}{\sqrt{2}\,\sigma_\xi}\right)\right]}
\def\oneplserf{\left[1+\erf\left(\frac{\xi(m,\eta^0(m-\mu(r)))-\xlim}{\sqrt{2}\,\sigma_\xi}\right)\right]}
\def\sigxi{\sigma_\xi}
\def\Axi{\cala_{\xi}}
\def\Axiinv{\cala_{\xi,{\rm inv}}}
\def\Az{\cala_z}
\def\Azinv{\cala_{z,{\rm inv}}}
\def\Axiprime{\cala'_{\xi}}
\def\Azprime{\cala'_z}
\def\betafac{\frac{\beta}{\sqrt{1+\beta^2}}}
\def\alfafac{\frac{\alpha}{\sqrt{1+\alpha^2}}}
\def\Abetafac{\frac{\Axi^2}{1+\beta^2}}
\def\Abfsqrt{\frac{\Axi}{\sqrt{1+\beta^2}}}
\def\Aalfafac{\frac{A_z^2}{1+\alpha^2}}
\def\Aafsqrt{\frac{A_z}{\sqrt{1+\alpha^2}}}
\def\atan{{\rm tan}^{-1}\,}
\def\dugc{\ifmmode D_{{\rm UGC}}\else$D_{{\rm UGC}}$\fi}
\def\logdugc{\ifmmode \log\dugc\else$\log\dugc$\fi}
\def\deso{\ifmmode D_{{\rm ESO}}\else$D_{{\rm ESO}}$\fi}
\def\logdeso{\ifmmode \log\deso\else$\log\deso$\fi}
\def\cint{C_{{\rm int}}}

\title{Homogeneous Velocity-Distance Data \\
for Peculiar Velocity Analysis. \\
III. The Mark III Catalog of Galaxy Peculiar Velocities}
\author{Jeffrey A.\ Willick$^a$, St\'ephane Courteau$^b$, S.M.\ Faber$^c$,
\\ David Burstein$^d$, Avishai Dekel$^e$, and Michael A.\ Strauss$^{f,g}$ \\
\bigskip
{\scriptsize
$^a$ Dept.\ of Physics, Stanford University, 
Stanford, CA 94305-4060 
{\tt (jeffw@perseus.stanford.edu)} \\
$^b$ NOAO/KPNO, 950 N.\ Cherry Ave., Tucson, AZ 85726 
{\tt (courteau@noao.edu)} \\
$^c$ UCO/Lick Observatory, University of California, Santa Cruz, CA 95064 
{\tt (faber@ucolick.org)} \\
$^d$ Arizona State University, Dept. of Physics and Astronomy, Box 871504,
Tempe, AZ 85287 
{\tt (burstein@samuri.la.asu.edu)} \\
$^e$ Racah Institute of Physics, The Hebrew University of Jerusalem,
Jerusalem 91904, Israel
{\tt (dekel@astro.huji.ac.il)} \\
$^f$ Dept.\ Astrophysical Sciences, Princeton University, 
Princeton, NJ 08544 
{\tt (strauss@astro.princeton.edu)} \\
$^g$ Alfred P.\ Sloan Foundation Fellow
}}
\begin{abstract}
This is the third in a series of papers in which
we assemble and analyze a homogeneous catalog of
peculiar velocity data. In Papers I and II, we
described the Tully-Fisher (TF) 
redshift-distance samples that constitute the
bulk of the catalog, and our methodology
for obtaining mutually consistent TF calibrations
for these samples. In this paper, we supply further
technical details of the treatment of the data,
and present a subset of the catalog in
tabular form. The full catalog, known as the
{\em Mark III Catalog of Galaxy Peculiar Velocities,}
is available
in accessible on-line databases, as described herein.
The electronic catalog incorporates not only the
TF samples discussed in Papers I and II, but also
elliptical galaxy \dnsigma\ samples originally presented
elsewhere. 
The relative zeropointing of the elliptical and spiral
data sets is discussed here.

The basic elements of the Mark III Catalog are the
observables for each object (redshift, magnitude,
velocity width, etc.) and inferred distances
derived from the TF or \dnsigma\ relations. Distances
obtained from both the forward and inverse TF 
relations are tabulated for the spirals. 
Malmquist bias-corrected distances are computed
for each catalog object using density
fields obtained from the \iras\ 1.2\ Jy redshift survey.
Distances for both individual
objects and groups are provided.
A variety of auxiliary data, including distances
and local densities predicted from the \iras\ redshift
survey reconstruction method, are tabulated as well.
We study the distributions of TF residuals for three
of our samples and conclude that they are well-approximated
as Gaussian. However, for the Mathewson \etal\ sample we demonstrate
a significant decrease in TF scatter with increasing velocity width.
We test for, but find no evidence of, a correlation
between TF residuals and galaxy morphology.
Finally, we derive transformations that map the apparent
magnitude and velocity width data for each spiral sample onto
a common system. This permits the application of analysis
methods which assume that a unique TF relation describes the
entire sample.
  
\end{abstract}


\section{Introduction}

Analyses of the peculiar velocity field in the local
universe can provide strong constraints on cosmological
models [cf.\ the
reviews by Dekel (1994) and
Strauss \& Willick (1995)]. Among other things, they hold the
promise of testing the gravitational instability mechanism
as the origin of large-scale structure, clarifying the relative
distribution of luminous and dark matter, and, when analyzed
jointly with full-sky redshift surveys, constraining
the value of the density parameter $\Omega_0.$ Detailed peculiar
velocity analyses require large samples of galaxies with both
redshifts and redshift-independent distance estimates. The latter
are notoriously difficult to obtain free of serious systematic errors.
It has been apparent for some time that a full
realization of the promise of peculiar velocity studies requires
redshift-distance catalogs sufficiently
large ($\simgt 10^3$ objects) as to minimize purely statistical errors, 
and prepared with great attention to uniformity so as to
minimize systematics.

In this paper, the third in a series, we present the first 
velocity-distance catalog to substantially meet these criteria.
In Papers I (Willick \etal\ 1995) and II (Willick \etal\ 1996),
we described the principles 
behind the catalog assembly and construction, and calibrated
the Tully-Fisher (TF) relations (Tully \& Fisher 1977)
for the individual spiral samples.
Here we address several issues that
were not dealt with in Papers I and II, and
present representative subsections of the final data set,
known as the {\em Mark III Catalog of Galaxy Peculiar Velocities.}
Because of its large size, the Mark III catalog is not presented
here in full, but has been
made available electronically
via on-line astronomical databases as described below (\S~6.4).
In later papers in this series (Faber \etal\ 1996,
Paper IV; Dekel \etal\ 1996, Paper V) we analyze 
the velocity field in the local universe derived from
the catalog. It is not our intention that 
the Mark III catalog remain the private domain of the present
authors. We hope, rather, that it will be widely exploited by members of 
the community.

The outline of this paper is as follows. In \S~2, we give
a broad overview of the principles behind the catalog's
construction, and clarify the nature of the 
redshift-independent distances it contains. 
In \S~3 we provide details
of the various corrections to which the TF observables (velocity
widths and apparent magnitudes) were subjected prior to use
in the TF relation. In \S~4 we 
tabulate the data used in the ``overlap comparison''
used to derive relative TF zero points between samples
(Paper II, \S~6).
Our method for
computing Malmquist bias corrections to sample galaxies
is described in \S~5, along with a discussion of further
subtleties of bias correction.
In \S~6, we first rederive 
inverse TF relation zero points (superseding
the inverse TF zero points derived in Paper II, \S~6),
and present the final forward and inverse TF relations
for the Mark III spiral samples.
We then present representative parts of the spirals catalog
and provide 
instructions for accessing the full catalog electronically.
The incorporation of the elliptical
galaxy sample of Faber \etal\ (1989) into the spiral database
is discussed in \S~7,
with special attention paid to 
the normalization of the elliptical and spiral distance scales.
In \S~8 we carry out a simple analysis of the TF residuals,
and test the usual assumption that they are Gaussian.
The motivation behind and procedure for putting the TF observables
for all samples on a common system characterized by a single
TF relation is presented in \S~9. We conclude the paper in \S~10
by briefly summarizing our procedures
and discussing various possible systematic errors which
might yet lurk in the catalog. 

\section{Overview of the Mark III Catalog}

Before giving the details of the data in the 
Mark III catalog, we provide a brief
overview of what the catalog contains---and what it does not. 
Our ultimate goal
is to construct a homogeneous database of redshift-independent distance
estimates for use in velocity field analyses such as 
POTENT (Dekel 1994). This pursuit is
in keeping with the approach of
Burstein in his electronic distribution of the Mark I (1987)
and Mark II (1989) catalogs. The
challenge is how to construct such a database from separate samples of galaxies
selected and observed in different ways by different observers. We have brought
together a disparate set of six spiral galaxy samples for which distance
estimates are obtained using the TF relation. The main properties
of these six spiral samples are summarized in Table~\ref{tab:summary};
full details of their selection criteria 
may be found in Papers I and II. 
Some of these samples (HMCL, MAT)
are based on $I$-band CCD photometry, some (W91CL, W91PP, CF) on $r$-band
CCD photometry\footnote{Although we treat W91CL and W91PP as distinct
samples (cf.\ Paper II, \S~3.1.2, for further explanation), photometrically
they are identical (Willick 1991). We will thus lump them together
at times when commenting on purely photometric
aspects of the data set, referring to them 
collectively as ``W91''.},
and one (A82) on $H$-band photoelectric photometry. Most are
based on \hi\ velocity widths, while one (CF) uses exclusively optical rotation
curves and one (MAT) a mixture of both \hi\ and optical widths. Furthermore,
the various samples typically probe different regions of the sky (maps of the
spatial distribution of these samples are presented by Kolatt \etal\ 1996).
To this already
disparate group of spiral samples, we are adding a 
sample of elliptical galaxies
(Faber 1989; distributed electronically 
by Burstein 1989 as the part of the Mark II Catalog)
whose distances are estimated using the \dnsigma\ relation. 

Because of this diversity of input data, our chief concern 
has been to ensure that the estimated galaxy distances are on a uniform system.
Papers I and II described how we sought to achieve this goal for the spiral
samples, but the overall
approach bears repeating here. We began with the assumption that the 
HMCL sample consisted of a uniformly measured
set of $I$-band apparent magnitudes $m$ and velocity width parameters
$\eta$ (see equation~\ref{eq:defeta} below). We further assumed
that the HMCL
clusters had vanishing radial peculiar velocities 
in the mean, which we justified
on the grounds of the sample's wide sky coverage and depth. This last
assumption enabled us to take the HMCL cluster radial velocities as
being, in the mean, fair measures of their
cosmological distances. 
Taken together,
our assumptions
enabled us to fit a single TF relation (zero point $A,$ slope $b,$
and scatter $\sigma$) 
to the entire HMCL sample. The zero point is such that the TF
relation yields distances 
in units of \kms. Such a distance is defined
as the part of the observed radial velocity due to the Hubble expansion alone. 
From this it follows
that the difference between the observed radial velocity and the TF
distance is a fair measure of the radial peculiar velocity 
(neglecting various bias effects; 
see below).

Our next step was to carry out analogous TF 
calibrations for the remaining spiral
samples, except that we did not initially 
assign final TF zero points.
Because these
samples are either not full-sky 
(W91CL, W91PP, CF, MAT) or very shallow
(A82), we argued that it was not safe to 
assume that their radial peculiar velocities
vanished on average (i.e., that redshift 
equals distance in the mean) and thus 
assign TF zero points as we had with HMCL. Instead, we relied on
an ``overlap procedure'' to establish the remaining 
zero points: We identified,
first, objects
in common between HMCL and W91CL and required that their TF distances
were the same on average, which determined the W91CL TF zero point.
We then did the same for W91PP, CF, MAT, and A82, in each case adjusting
the TF zero point to obtain consistent distances for 
objects  in common with all already-calibrated samples (cf.\ Paper II, \S~6).
In this way, we
argued, the distances derived from the various samples were guaranteed to be
on a uniform system. 

Several other aspects of the approach developed in Papers
I and II bear reemphasis as well. 
First, we adopt
the raw measurements (apparent magnitudes and velocity widths) reported
by the original authors, but subject these 
quantities to our own, uniform correction
procedures (detailed below in \S~2). By doing this we ensure that
spurious differences between samples are not introduced as a result of the
distinct approaches to raw data correction present in the original papers.
Second, the TF relations of the various samples are calibrated with careful
attention paid to the role of {\em selection bias\/} 
(Willick 1994), specifically,
the effects of magnitude, diameter, and other limits that define
the data sets. 
In order to make the selection bias corrections,
we have devoted considerable effort 
to characterizing sample selection criteria
as quantitatively as possible. 
Selection bias is especially strong when the {\em forward\/}
form of the TF relation---absolute magnitude considered as a function of
velocity width---is employed. 
Such bias is weak or negligible, however, when the {\em inverse\/}
form of the relation---velocity width considered as a function of absolute
magnitude---is used. In Papers I and II, 
we calibrated both forward and
inverse TF relations for each sample. The latter form of the relation
is characterized by an inverse slope $e$ and inverse zero point $D,$
which are not trivially related to their forward counterparts
(i.e. $e\neq b^{-1},$ $D\neq A$; cf.\ Appendix C for further discussion). 
Relative distances for groups or clusters
resulted from both the forward and inverse TF calibrations. 
The large corrections for forward TF selection bias were
validated by demonstrating good agreement
between the forward and (nearly unbiased) 
inverse distance moduli of these groups or clusters 
(cf.\ Paper I, \S~5 and Paper II, \S~2.2.7, \S~3.1.5, \S~5.2.6).\footnote{In
view of the nearly unbiased nature of the inverse relation one can ask
why it is worthwhile working with the forward relation at all. The answer is
that in Method I velocity field analyses (cf.\ \S~2.1) such as POTENT, the forward relation
yields distances with relatively straightforward Malmquist bias corrections
that are independent of sample selection. Inverse TF distances used in
a Method I analysis require Malmquist corrections that depend on both
sample selection criteria and the luminosity function.
See Strauss \& Willick (1995), \S~6.5, for futher details.} 

We have recently recognized, however, that our
assignment of final inverse TF zero points in Paper II did
not lead to consistent forward and inverse group distances
within each sample.
We describe this problem in
greater detail in \S~6.1, and discuss the method
we have adopted for rederiving final inverse TF zero points. 
The new zero points differ from the old (cf.\ Paper II, Table~12) 
at the level of $\sim 0.05$ mag. 
Final forward and inverse TF parameters for all the Mark III
spiral samples are presented in \S~6.1.
None of the important conclusions
of Papers I and II are affected in any way by 
this revision in our procedure.
In particular, the validation of the 
forward bias corrections by comparison of forward and inverse
distance moduli did not depend on final TF zero points. 

\subsection{TF Distances in the Mark III Catalog}
The procedures just described yield 
fully corrected
TF observables ($m,\eta$) for each object, as well
as forward and inverse TF parameters
(zero point, slope, and scatter) for each sample (cf.\ 
Table~\ref{tab:finaltf}). From these data we
may derive any number of redshift-independent
distance estimates for individual galaxies. The ones
we actually tabulate in the Mark III spiral singles catalog
are the following:
\begin{enumerate}
\item A {\em raw forward TF distance,} 
$d_{{\rm TF}}=10^{0.2[m-(A-b\eta)]}.$ These are the
distances which were used (Paper II, \S~6) to bring
the spiral samples onto a uniform TF distance scale.
Such distances are not, however, suitable as input
directly into velocity analysis methods: they
are strongly affected by Malmquist bias or
selection bias, depending on whether a {\em Method I\/}
(TF-distance taken as the \apriori\
distance indicator)
or {\em Method II\/} (redshift taken as the
\apriori\ distance indicator)
approach is taken (cf.\ Strauss \& Willick 1995,
\S~6.4.1, for further explanation).
\item A {\em raw inverse TF distance,}
$d_{{\rm TF}}^{{\rm inv}}=10^{0.2[m-(D-\eta/e)]}.$ 
Such distances are not suitable for a straightforward
Method I analysis, but are relatively unbiased
in a Method II analysis. For reasons described
in \S~6.1, the raw inverse distances do not
necessarily agree in the mean with their forward counterparts.
\item A {\em Malmquist-corrected forward TF distance,} 
$d_{{\rm TF}}^{{\rm mc}}.$ Computation of this quantity
is discussed in \S~5. 
In general, this is the distance that should
be used in a Method I velocity analysis, and is the
quantity used in POTENT, subject
to the caveats discussed in \S~5.1. 
\end{enumerate}
In the spiral groups catalog we provide two
measures of distance for the clusters of Paper I
and the field galaxy groups of Paper II: selection-bias
corrected forward and inverse TF distances.  
In contrast to the forward and inverse TF distances
to individual galaxies, the group distances
agree, by construction, in the mean (\S~6.1).
These group distances may be used as they stand
in a Method II analysis. They will be subject to
a subtle though diminished Malmquist bias
in a Method I approach, as we discuss further below.

\subsection{Further Discussion}
It is important for users of the catalog 
to bear in mind three
caveats about the TF distances contained therein. 
First, {\em which of the various
measures of TF distance to use depends on the method of velocity field analysis
employed.} For example, while a Malmquist-corrected forward distance is
generally appropriate for a Method I analysis, it is incorrect to use such a
distance in a Method II analysis, in which redshift-space information is
used as the \apriori\ distance indicator. Second, {\em we have not included
all possible measures of TF distance in the catalog.} For example, we do not
calculate a Malmquist-corrected inverse TF distance, which has
properties quite distinct from its forward counterpart; we will
address this issue in a future paper 
(Eldar, Dekel, \& Willick 1997; see Strauss \&
Willick 1995, \S~6.5.5 for further discussion). 
Third, {\em the refined distance
estimates we do tabulate are based on certain model-dependent assumptions
and are not necessarily correct in an absolute sense.} 
As we discuss more fully in \S~5, our 
Malmquist bias corrections
are based on an assumed model of the underlying galaxy density field. 
The selection-bias corrected group distances depend on
the validity of our quantitative model of sample selection
criteria (although for the inverse TF relation the
dependence is small).
One should critically examine all such model dependencies 
when interpreting velocity field analyses based
on the Mark III---or, indeed, any other redshift-distance---catalog.

Two issues implicit in these caveats
merit further comment. First, we have
neglected a potentially significant effect 
in computing the Malmquist-corrected
forward TF distances that appear in the catalog: 
the role of a redshift limit
in the definition of a velocity distance sample. 
If sample objects are required to lie within some maximum redshift,
then in the vicinity of that limit the proper Malmquist 
correction can differ considerably
from the usual expression (e.g., equation~\ref{eq:erD}
below), which
assumes that objects may lie at any 
distance along the line of sight. In \S~5.1,
we discuss this problem in some detail, and indicate how the effect may be
accounted for in a given analysis (and how, in fact, 
it is done in recent implementations
of POTENT). However, as discussed
in \S~5.1, the redshift limit effect is unimportant
for most Mark III galaxies. Moreover, accounting
for its effect is quite model-dependent (\S~5.1).
Consequently, we neglect redshift
limits in computing the
Malmquist corrected distances in the catalog,
but provide sufficient information for the
user to take them into account if desired.

The second
issue concerns the Malmquist corrections that should be applied to groups.
As already noted, we tabulate selection-bias corrected group distances
in the catalog. These distances are the correct ones 
to use in Method II analyses.
However, one may also use such groups in Method I analyses such as
POTENT. In that case, 
the selection-bias corrected group distances
play a role roughly analogous to the raw individual 
galaxy distances in an ungrouped analysis,
but with smaller distance errors. One might infer from 
this that the corresponding
Malmquist correction is a straightforward adaptation of
the singles formula. 
However, this is not the case;
now, in addition to the standard Malmquist effects (density and volume) that
affect the probability of selection as a function of distance along the line of
sight, there is also the effect of the relative likelihood that an object is in
a group, or is single, as a function of distance. We have recognized this
effect for several years and incorporated a correction for 
it into preliminary POTENT
analyses (e.g., Dekel 1994; Hudson \etal\ 1995). 
However, our understanding and treatment of
this effect 
are still being refined; recent
work with the simulated catalogs of Kolatt \etal\ (1996) has suggested that
our initial approach to the problem may require modification. Because this subject
is in flux, we have elected to present only selection-bias corrected group and
cluster distances. We hope to present more definitive conclusions on this subject
in the future.

In summary, we have chosen to present only raw 
(forward and inverse) TF distances,
and those processed measures of TF distance (forward Malmquist corrected
for singles,
forward and inverse selection-bias corrected for groups) 
whose computation is straightforward
and based on reasonably well-founded assumptions. We have neglected several
effects (redshift limit, grouped versus ungrouped fraction) whose 
proper correction may be 
ambiguous or model-dependent. We have attempted in
this discussion to clarify these points.
What must be borne in mind above all, however, is that all refinements of
the redshift-independent distances
will be to no avail, or even lead to spurious results, if the user
of the catalog does not keep in mind a fundamental
tenet: {\em the proper measure of TF distance depends on the type of analysis
adopted.} Indeed, for some analytic approaches one takes the TF observables
($m,\eta$) as the basic input quantities and bypasses the distances altogether.
A detailed discussion of these issues is provided by
Strauss \& Willick (1995, \S~6.5) and references therein.

To the discussion above we 
add an important if perhaps obvious remark.
The scientific analysis of any catalog is only as good as the data it contains.
We believe that our procedures for producing the catalog are valid and that the
results are reliable. We cannot exclude the possibility, however, that we have
erred in some of the basic assumptions that underlie the catalog construction.
We have already emphasized in Papers I and II that the global TF zero point 
could be in error were the HMCL clusters to possess a net
radial peculiar velocity. 
A more serious possibility is that the HMCL sample
could be less uniform with respect to its Northern and Southern hemisphere
components than we have assumed. Because HMCL is the glue that holds the
Mark III spirals together, any such nonuniformity would propagate throughout
the data set. The best way to test for such possibilities is to continue to
subject the Mark III catalog to cross-checks with new data as it comes in.
Plans are presently underway for such checks, and, as we reiterate in this
paper's conclusion, we will seek to keep the community apprised of the
outcome of this program.

\section{Corrections to the Observables}
\label{sec:corrsect}

The TF relation is applied to {\em corrected,} rather
than raw, values of the input data, namely
apparent magnitudes and linewidths.
The corrections are for
effects such as extinction and inclination which affect
the values of the observables but are of no fundamental
relevance to the scientific analysis. 
Because these corrections
can be sizable, they must be considered as hidden
but nonetheless integral parts of the TF calibration.
A change in
the details of the corrections would entail changes
in the TF relations themselves. Before we describe the corrections
in detail, some general remarks concerning our approach
are in order. 

We have adopted a uniform set of rules
for the corrections to the observables,
as this contributes to the homogeneity of the samples.
These rules are, in general, not the same as those adopted 
by the original authors of each Mark III sample. 
Consequently, the values of the TF observables
found in the Mark III catalog differ
from those originally published. 
At the same time, though, we have attempted to minimize
these changes by departing to the
smallest degree possible
from the approach of
the original authors, consistent with the
requirement of uniformity.
For example, MAT 
used a different algorithm for computing
\hi\ velocity widths than did the other samples
based on 21 cm linewidths. 
The MAT widths are thus quite different from
those in other samples for the same objects. 
We do not attempt to force the MAT widths onto
the system used by the other samples; instead, the
difference is accounted for by the distinctly different
TF slope found for MAT as compared with, say, HMCL\footnote{The
exception to this procedure is when we place the observable data
onto a common system in \S~8.}.
Another feature of our approach is that we forgo corrections
to the observables that depend specifically on
morphological type, for two reasons. 
First, we have not found that morphological information
correlates in any way with residuals from
the Tully-Fisher relation, as we demonstrate below (\S~8).
Second, in many cases
the existing imaging data not allow
us to assign a reliable morphological type.
This is particularly
true of the many sample objects which are
relatively distant ($\!\simgt 5000\ \kms$), as well as 
objects viewed at high inclination angles.

\subsection{Details of the Corrections}

There are four important corrections which we
make to the observables: an inclination correction
applied to velocity widths, Galactic and internal
extinction corrections applied to the apparent 
magnitudes, and a cosmological correction applied
to the magnitudes.   

\subsubsection{Inclination correction}

Velocity widths must be corrected for projection. If we
begin with a velocity width $\Delta v$ corrected only
for redshift (\ie, the raw width divided by $(1+z)$),
the width corrected for projection is
\begin{equation}
\Delta v^{(c)} = \frac{\Delta v}{\sin i}\,,
\label{eq:dvcorr}
\end{equation}
where $i$ is the inclination of the galaxy to the
line of sight. We remind the reader that the TF relation
is expressed not directly in terms of $\Delta v^{(c)},$
but rather in terms of the {\em velocity width parameter} 
\begin{equation}
\eta \equiv \log\Delta v^{(c)} -2.5,
\label{eq:defeta}
\end{equation}
with $\Delta v^{(c)}$ expressed in \kms.

We calculate the inclination angle
$i$ in all cases according to the formula
\begin{equation}
\cos^2 i = \left\{ \begin{array}{ll}
\frac{(1-\varepsilon)^2-(1-\varepsilon_{max})^2}{1-(1-\varepsilon_{max})^2}\,, &
\varepsilon < \varepsilon_{max}\,; \\
0\,, & \varepsilon \geq \varepsilon_{max}\,.
                   \end{array} \right.
\label{eq:incleps}
\end{equation}
In the above equation, $\varepsilon$ is the ellipticity of
the image of the galaxy disk, and $\varepsilon_{max}$ is the ellipticity 
above which the galaxy is automatically assigned an
inclination of $90\degs,$ i.e., the typical ellipticity 
of a galaxy seen edge-on.
While the inclination formula~(\ref{eq:incleps}) is a standard one
(e.g., Bothun \etal\ 1985),
some workers (e.g., Aaronson, Huchra, \& Mould 1979) have adopted
modifications of it in TF work, while others have taken
$\varepsilon_{max}$ to have
a morphological type dependence. We apply equation~(\ref{eq:incleps})
in all cases without modification. For three of
our samples (A82, MAT, and HMCL) we use
the value $\varepsilon_{max}=0.80.$ However, for the $r$-band
samples (W91 and CF), we use $\varepsilon_{max}=0.82$. 
This difference is trivial, and is made
only for consistency with the original authors.
For the samples (HMCL, W91, CF, and MAT) based on
CCD photometry, we use the ellipticities determined
by the original authors from the CCD images. For the
one sample based on $H$-band aperture photometry (A82),
we compute ellipticities from the blue axial ratios given
in the RC3 Catalog (de Vaucouleurs \etal\ 1991), following
Tormen \& Burstein (1995).

\subsubsection{Galactic Extinction Correction}
The Galactic extinction correction 
is taken to be proportional 
to the Burstein-Heiles (Burstein \& Heiles 1978, 1984; BH)
reddening estimate in the direction of each galaxy.
If we write the BH reddening
estimate as $E(B-V)$, then Galactic extinction
correction for bandpass $j$ is given by
\begin{equation}
A_G(j) = f_B(j) A_B = f_B(j) \times 4 E(B-V)\,,
\label{eq:defag}
\end{equation}
where $f_B(j)$ is the ratio of extinction
in bandpass $j$ to that in $B,$
and we assume that the $B$-band extinction is four
times the $(B-V)$ reddening.\footnote{This assumption is
not made universally; in particular, the RC3 catalog assumes 
that $A_B=4.3 E(B-V)$.} We have adopted the following
values for $f_B$ for the various samples:
$f_B=0.10$ for the $H$ bandpass (A82);
$f_B=0.42$ for the $I$ bandpass (HMCL and MAT);
and $f_B=0.56$ for the $r$ bandpass (W91, CF). The derivation of
this last value, 
for a somewhat nonstandard bandpass, 
is discussed by Courteau (1992).

\subsubsection{Internal Extinction Corrections}

\def\calrmin{\calr_{{\rm min}}}
\def\calrmax{\calr_{{\rm max}}}
\def\cint{C_{{\rm int}}}
A number of possible internal extinction formulae
exist, as summarized by Willick (1991).
However, it is difficult
to distinguish between the quality of the various
formulae; most can adequately describe
the data provided the right parameters
are chosen. As discussed in Paper I, \S~2, we  
adopt one of the simpler
forms. We write
the logarithm of the (major to minor) axial ratio
by $\calr$, \ie,
\begin{equation}
\calr = -\log (1-\varepsilon)
\label{eq:defcalr}
\end{equation}
where $\varepsilon$ is, as before, the apparent ellipticity
of the galaxy disk. We then 
compute the internal extinction correction in bandpass $j$ as
\begin{equation}
A_{{\rm int}}^j(\calr) = \cint^j\times\left\{ \begin{array}{lll}
		   \calrmin - \calr_0^j\,, & \calr < \calrmin\,;\\
		   \calr - \calr_0^j\,, & \calrmin \leq \calr \leq \calrmax\,;\\
		   \calrmax - \calr_0^j\,, & \calr > \calrmax\,.
		   \end{array} \right.
\label{eq:defintcorr}
\end{equation}
$\cint^j$ is the bandpass-dependent internal extinction coefficient.
As discussed at length in Papers I and II, we determined its
value for the various bandpasses by minimizing TF scatter.
We found, in particular, $\cint^r=\cint^I=0.95,$ and $\cint^H=-0.30$
(see next paragraph for further discussion).\footnote{Bottinelli
\etal\ (1995) have also addressed
the issue of internal extinction using minimization of TF
residuals. They worked with $B$-band photometric data
and found $\cint^B=1.67\pm 0.23.$ This is larger than
what we find for the $r$ and $I$ bandpasses, as is expected
for shorter wavelength photometry. While these results
are reasonably consistent, quantitative agreement
is difficult to establish in the absence of a satisfactory
theory of internal extinction in galaxies.}
The quantity $\calr_0^j$ that appears in equation~(\ref{eq:defintcorr})
is the value of the logarithmic axial ratio 
to which the internal extinction correction
is referenced. We have used
$\calr_0=0$ (correction to
face-on orientation) for the $H$-band (A82) and $I$-band (HMCL, MAT)
samples, and $\calr_0=0.418$ 
(correction to \sm 70\degs\ inclination) for the $r$-band 
(W91, CF) samples. The latter value is
adopted for consistency with the original
authors, who preferred to keep the
absolute size of the 
correction small. It should be clear
that a non-zero value of $\calr_0$
has no physical
significance whatsoever, as it is ultimately 
absorbed into the TF zero point.
The quantities $\calrmin$
and $\calrmax$ in equation~(\ref{eq:defintcorr})
reflect the ``saturation'' of the internal extinction
effect at low and high inclination. We have adopted
the values $\calrmin=0.27$ and $\calrmax=0.70$ for
all the spiral samples. These values were arrived at
by adjusting them until
TF residuals at high and low axial ratios exhibited
no trends. 
In Paper II, we considered the possibility that internal extinction
is luminosity-dependent, as has recently been suggested by Giovanelli
(1995). We carried out careful tests for 
such an effect but found no evidence for it (Paper II, \S~2.3.1
and \S~3.2.1).

Two aspects of the derived internal extinction coefficients warrant
further comment. First, we have recognized a significant (though
harmless) error in our estimates of the uncertainties in
these coefficients in Papers I (\S~3.2.3, \S~4.2.1) and II
(\S~2.3, \S~3.2, \S~5.2.7).  Specifically, we based those
estimates on an erroneous statement (Paper I, 3.2.3) 
of the relationship between 
a $\chi^2$ statistic for the TF calibration fit and the TF $\sigma.$ 
The correct statement is $\chi^2(\cint)\simeq N_{{\rm eff}}\,
\sigma^2(\cint)/\sigma_{{\rm min}}^2,$ where $\sigma^2(\cint)$
is the TF scatter for an arbitrary value of $\cint,$ and
$\sigma_{{\rm min}}^2$ is the TF scatter for the value
of $\cint$ that minimizes scatter. With this corrected expression
for $\chi^2(\cint)$, one can once again go through our basic argument
that a 65\% confidence interval on
$\cint$ is obtained by asking, for what values of $\cint$ does
$\chi^2$ change by 1 unit from its minimum value of $N_{{\rm eff}}.$
The result is that our confidence intervals on $\cint$ for all three
bandpasses ($I,$ $r,$ and $H$) were too wide. In particular,
our final estimate of $\cint^I$ is uncertain by  $\simlt 0.1,$
and of $\cint^r$ and $\cint^H$
by $\simlt 0.15,$ 
roughly half as large as the uncertainty estimates given in
Paper II, \S~8. (The reduction is comparatively modest,
despite the fact that our $\chi^2$s were badly off, because of
the flat behavior of $\chi^2$ near its minimum.)

Second, we emphasize that the negative coefficient of
$H$-band internal extinction does not
represent an important physical distinction between $1.6\mu$
and far-red optical galaxian light. It reflects, rather, the
fact that the original aperture photometry from which the
$H$-band magnitudes were derived has been scaled to standard
{\em diameter\/} measurements. These diameters were
not corrected for inclination (Tormen \& Burstein 1995).
Thus, any systematic dependence of diameter on inclination
would manifest itself as an inclination dependence of
the apparent magnitudes as well. (The CCD-measured magnitudes
of HMCL, W91, CF, and MAT are, in contrast, {\em total\/} magnitudes and 
thus unaffected by diameter measurements.) In particular, if
galaxies actually get systematically larger with increasing
inclination, as is certainly possible, then the corresponding $H$-band
magnitudes would get systematically brighter.
This is the sense of the effect we have detected, and
is the most likely explanation of the negative value of $\cint^H.$

\subsubsection{Cosmological Correction}

The final correction we apply to the apparent magnitudes
is for cosmological effects (by which we mean all effects
associated with redshift; see below).  
It is closely related but not identical
to the standard K-correction (\eg, Oke \& Sandage 1968;
Pence 1976).
In many previous studies (\eg, Mathewson \etal\ 1992), 
this K-correction has been applied to CCD total magnitudes
used for TF purposes.
However, such a procedure is not rigorously correct.
In what follows, we will discuss why this is,
and derive a cosmological correction appropriate for
CCD magnitudes used as input to the TF relation in
peculiar velocity analyses. Although the practical differences
from earlier work are small in the present application, our
modification may be significant in studies that
apply the TF relation to galaxies at redshifts $\simgt 0.1.$

Before proceeding, a clarification is desirable.
We use the term ``cosmological correction'' in the same
sense that Oke \& Sandage (1968) use  
``K-correction'': to signify correction for the
effects of both the shift of the spectrum relative
to the observational bandpass and the change in
spacetime geometry with increasing redshift. 
Our cosmological correction differs from the
K-correction for two reasons,
the first quite straightforward and the second considerably
more subtle. First, standard K-corrections (e.g., Oke and
Sandage 1968; Pence 1976) are
derived under the assumption that it is the {\em energy\/}
detected from the source
that determines apparent magnitude. However, with
CCD photometry it is instead the number of photons
detected (as recognized by Schneider, Gunn, \& Hoessel 1983).  
This difference must be accounted for both in the way the
spectral shape is characterized (as we do in this section) and
in the mathematical derivation of the correction (as we
do in Appendix A). 
Second, the ``distance''
one wishes to obtain from comparison of observed apparent
magnitudes and inferred absolute magnitudes in a
velocity field analysis is not
one of the standard measures of cosmological distance
(e.g., the angular diameter or luminosity distances).
Instead, it is the quantity $cz_c,$ where
$z_c$ is {\em the
redshift the object would possess if its peculiar
velocity were zero.} 
It is
this particular measure of cosmological distance
which, when compared with the observed
redshift $cz,$ yields the radial component of peculiar velocity.
  
The K-correction $K(z)$ (\eg, Oke \& Sandage 1968) 
is defined so that 
\begin{equation}
m - K(z) - M = 5\log d_L\,,
\label{eq:defdL}
\end{equation}
where $m$ is the observed apparent magnitude, $M$ is
the absolute magnitude of the standard candle, and the
{\em luminosity distance}  is given by (cf.\ Weinberg 1972) 
\begin{equation}
d_L(z) = \frac{c}{q_0^2} \left[zq_0 + (q_0 - 1)(-1+\sqrt{2q_0z+1})\right]\,.
\label{eq:dLofz}
\end{equation}
(In equations~\ref{eq:defdL} and~\ref{eq:dLofz} we have conformed
to our usual practice of defining distance in \kms\ units and
taken $1\ \kms$ as the fiducial distance at which absolute
magnitude is defined.) 
\def\ktf{K_{{\rm TF}}}
By contrast, for the purposes of peculiar velocity
analysis, the corrected distance
modulus should correspond to the distance $r=cz_c,$
where 
$z_c$ is the ``cosmological redshift'' defined
in the previous paragraph.
Thus, the relevant cosmological correction for our
purposes, $\ktf,$ is defined by the relation\footnote{In 
equation~(\ref{eq:defktf}),
$\ktf$ is written as a function of both $z$ (the observed,
or heliocentric, redshift) and $z_c,$ as it is the former
that determines the amount by which the spectrum is shifted,
while it is the latter that determines specifically cosmological
effects (see Appendix A for further details).
The distinction is of course very small, but we
preserve it in our analysis procedure, as discussed further in the text.}
\begin{equation}
m - \ktf(z,z_c) - M(\eta) = 5\log cz_c\,.
\label{eq:defktf}
\end{equation}
Because $d_L \neq cz_c,$ as we show in Appendix A, 
the K-correction defined by equation~(\ref{eq:defdL}) is
not equal to $\ktf(z,z_c)$ as defined by equation~(\ref{eq:defktf}).
In particular, while $K(z)$ is independent
of $q_0$---for which $d_L$ may be used
as a diagnostic---$\ktf(z,z_c)$ is not.
The distinction between the classical K-correction
and the cosmological correction
suitable for peculiar velocity work
was first recognized by Lynden-Bell \etal\ (1988),
who considered peculiar velocities
estimated from the \dnsigma\ relation. The corresponding
expression that applies for the TF case is different.
The full derivation of $\ktf$
is outside our main line of argument, and is presented
in Appendix A. The result (to first order in $z$ and $z_c$) is 
\begin{equation}
\ktf(z,z_c) = 1.086\,[(\epsilon + 2)z - (1+q_0)z_c]\,,
\label{eq:mykcorr}
\end{equation}
where $\epsilon$ is the power-law index (which we
model as depending on the velocity width parameter $\eta,$
as discussed below) 
of the photon number distribution $N(\lambda)$ (see Appendix A).
It is the photon number, rather than the energy flux, distribution
that is relevant for CCD magnitudes, as discussed above.

In applying equation~(\ref{eq:mykcorr}) to the Mark III TF
samples, 
we take $z$ to be the heliocentric redshift, and
estimate $z_c$ by the cosmic microwave background
frame (CMB) redshift. 
While peculiar velocities might invalidate
this estimate in any given instance, 
we expect that on average
the CMB redshift is a good estimator of the
cosmological redshift. As discussed above, 
equation~(\ref{eq:mykcorr}) also  
contains the deceleration parameter $q_0$, for which 
we must thus adopt a value.
We take $q_0=0.25$,
halfway between an open and flat universe.
It is important to bear in mind that although we
are obliged to make these uncertain assumptions ($z_c\simeq z_{{\rm CMB}}$
and $q_0=0.25$), the effect on our data analysis by adopting
plausible alternatives 
would be inconsequential, given that the mean
redshift of the sample is $\simlt 4000\ \kms.$

\begin{figure}[th]
\centerline{\epsfxsize=4.0 in \epsfbox{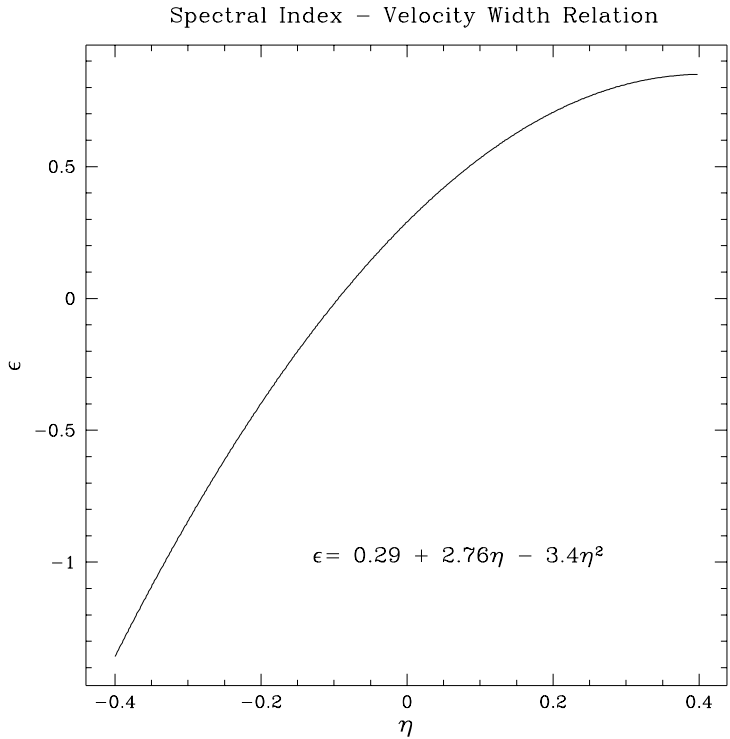}}
\caption{The relation between the
photon number spectrum power law index $\epsilon,$
and the velocity width parameter $\eta,$
adopted for the cosmological correction applied
to apparent magnitudes for the $r$ and $I$ band
Mark III spiral samples.}
\label{fig:eps_eta}
\end{figure}

The power-law exponent $\epsilon$ in equation~(\ref{eq:mykcorr})
must be properly modeled in order to avoid 
systematic errors. 
In previous work (\eg, Han 1991; Mathewson \etal\ 1992),
this effect has been approximated by assuming 
spectrum shape to be a function of morphological type.
As noted above, we consider morphological
information to be of limited accuracy 
for objects distant enough
that the cosmological correction is meaningful. 
We thus
adopt the following alternative criterion of spectral shape.
As shown by Willick (1991), the $r-I$ colors of
spiral galaxies 
correlate well with their 
velocity widths. The colors are
in turn a measure of the spectrum shape;
Willick (1991) calibrated the latter
effect by fitting the photon number power law
indices of spectrophotometric
standard stars to their $(r-I)$ colors.
Combining the color-velocity width
and spectrum shape-color
correlations, Willick (1991) derived
the $\epsilon$--$\eta$ relation
for spiral galaxies shown in Figure~\ref{fig:eps_eta};
the explicit formula for $\epsilon(\eta)$ is given in the plot.
The sense of the relation is that relatively 
faint ($\eta<0$) spirals tend to be blue ($\epsilon < 0$),
while luminous spirals tend to be red.
The trend saturates for the most luminous galaxies,
however. The variation in $\epsilon$ over the
range of observed width parameters is such that
the cosmological correction at $z\simgt 0.02$ can
differ by several hundredths of a magnitude for bright
as compared with faint galaxies. This corresponds
to distance differences of $\sim 100\ \kms,$ and
is not negligible.

Finally, then, we apply the following
$\eta$-dependent redshift correction
to apparent magnitudes in the samples based on
$r$- and $I$-band CCD photometry
(HMCL, W91, CF, MAT):
\begin{equation}
\ktf(\zhel,z_{{\rm CMB}},\eta) = 1.086\,[(\epsilon(\eta) + 2)\zhel - 1.25 z_{{\rm CMB}}]\,,
\label{eq:kcorrprac}
\end{equation}
where $\epsilon(\eta)$ is given in Figure~\ref{fig:eps_eta}.
This correction is {\em not} applied to the one sample (A82)
which uses $H$-band aperture magnitudes, for two reasons.
First, the $\epsilon(\eta)$ relation derived
by Willick (1991) does not apply at $H$. Second and more importantly,
the $H$-band magnitudes are tied to photographic diameters
(Tormen \& Burstein 1995),
which display a rather different behavior with increasing
redshift. As mentioned in Paper II,
for A82, we instead apply the simple redshift correction
$K(z)=1.9\zhel$ derived by Aaronson \etal\ (1980).
As A82 objects lie 
mainly at redshifts $\!\simlt 2000\ \kms$, their
cosmological corrections are very small
in any case.

\subsubsection{Summary}

We have standardized the corrections to
the raw observables for the six spiral samples. A
single formula, equation~(\ref{eq:incleps}), 
is used to compute inclinations and
thus deprojected velocity widths. The raw apparent magnitudes
undergo corrections for Galactic and internal extinction,
and for cosmological effects. If we denote by $m$
the corrected apparent magnitude, and by $m_j$ the
raw apparent magnitude (where $j=r$ or $I$), then
our full correction procedure is described by
\begin{equation}
m = m_j - A_G(j) - A_{{\rm int}}(j,\calr) - 
\ktf(\zhel,z_{{\rm CMB}},\eta)\,,
\label{eq:magcorr}
\end{equation}
where $A_G$ is given by equation~(\ref{eq:defag}),
$A_{{\rm int}}$ by equation~(\ref{eq:defintcorr}), and
$\ktf(\zhel,z_{{\rm CMB}},\eta)$ by equation~(\ref{eq:kcorrprac}).
For the $H$-bandpass, the equation is the same, except that
the cosmological correction is replaced by the
simple expression $1.9\zhel,$ as discussed above.

\section{Presentation of the Overlap-Comparison Data}

The reliability of the Mark III data for probing
large-scale peculiar motions depends 
on our ability to tie together the various samples
in a uniform way.
As discussed in Paper II, \S~6, 
we have done this by first identifying 
galaxies present in two or more of the Mark III samples
(``overlap'' objects), and then determining
relative TF zero points 
by minimizing TF distance modulus
differences (Paper II, \S~6).  The global TF zero point
was set by the HMCL sample (cf.\ Paper I, \S~3.2.2).\footnote{The 
elliptical data
cannot of course be normalized to the spirals via this procedure.
In \S~7 we discuss our method for establishing
the elliptical distance scale.}

Because the sample-to-sample matching is such an
important part of our procedure, we present in
Table~\ref{tab:match} the complete 
list of the 403 individual galaxies which participate in
the overlap comparison. 
The objects are
listed in order of ascending heliocentric
redshift. Column~1 lists
the Principal Galaxy Catalog (Paturel \etal\ 1985; PGC) number of the
object. This number provides a unique way of identifying
the galaxy. (Its common name or names may be found
by cross-referencing the PGC number with Table~3,
which lists all Mark III galaxies.) Columns~2 and~3
list the Galactic longitude ($l$) and latitude
($b$) in degrees. Column 4 lists the heliocentric
redshift of the object in \kms, averaged over
the two or more samples in which the object appears.
In the great majority of cases, the individual redshift
measurements agree to within $<100\ \kms.$ 
Redshift differences greater than this were found
for only six of the overlap objects. In these instances,
we used the value deemed most reliable for all samples. 

Columns~5--22 list the TF observables $(m,\eta),$ and the
raw forward TF distance $d_{{\rm TF}},$ 
for each of the Mark III samples in which the
object is found. This TF distance is not corrected
for Malmquist or selection bias and is expressed
in \kms\ units. Columns~5--7 list the data for
HMCL; columns~8--10 for W91CL;
columns~11--13 for W91PP; columns~14--16
for CF; columns~17--19 for MAT;
and columns~20--22 for A82.
If the object in question does not appear in a particular
sample, all three values ($\eta,$ $m,$ and $d_{{\rm TF}}$)
are simply listed as zero. 

Twenty HMCL galaxies listed in Table~\ref{tab:match}, indicated
by a superscript $a,$ are found in the HMPP subset of HMCL,
but not in the restricted HMCL sample used in the analyses
presented in Papers I and II, which excluded
the HMPP subset. However, these 20 objects 
are used in the common system definition presented in
\S~8. There are, in addition, six objects which appear
in both the HMPP subset and in the restricted HMCL sample
of Papers I and II. The HMPP data for these galaxies
do not appear in Table~\ref{tab:match}, but may
be found in Table~3. These objects are NGC 444
(PGC 4561); NGC 452 (PGC 4596); UGC 841 (PGC 4735);
UGC 987 (PGC 5284); NGC 536 (PGC 5344); and UGC 1066 (PGC 5563).

\section{Malmquist Bias Corrections}

When TF or \dnsigma\ distances are used in a {\em forward,}
{\em Method I\/} analysis (cf. Strauss \& Willick 1995, \S~6.4),
they must be corrected for Malmquist bias in order to yield
unbiased peculiar velocities. Malmquist bias arises because
objects with a given TF-inferred distance lie in reality
at a range of true distances because of TF errors. 
The average true distance of a set of objects with
a given TF-inferred distance depends on the underlying
galaxy density field as well as on the TF magnitude scatter $\sigma.$
A variety of approaches
to Malmquist correction are possible (Lynden-Bell \etal\ 1988;
Willick 1991; Landy \& Szalay 1992; Dekel 1994; Hudson 1994; Hudson \etal\ 1995;
Freudling \etal\ 1996). Our technique will follow
that outlined by Strauss \& Willick (1995).

The main complication is that the
Malmquist bias correction to a given galaxy depends nonlocally on
the underlying galaxy density field. 
In particular, if 
$d=10^{0.2(m-M(\eta))}$ is the raw forward TF
distance, then 
the expected {\rm true\/} distance $r$ 
is given by (Strauss \& Willick, \S~6.5.2).
\begin{equation}
E(r|d) = \frac{
\int_0^\infty r^3 n(r) \explnrD\,dr}{\int_0^\infty r^2 n(r) \explnrD\,dr}\,,
\label{eq:erD}
\end{equation}
where $n(r)$ is the (real-space) galaxy number density along the
line of sight, and
\begin{equation}
\Delta \, \equiv \, \left(\frac{\ln 10}{5}\right) \sigma \, \simeq \, 0.46\, 
\sigma
\label{eq:defDelta}
\end{equation}
is the fractional distance estimation error 
due to the TF magnitude scatter $\sigma.$
If $n(r)$ were effectively constant on the scale
($\sim \Delta d$) of TF distance errors, the above expression
would reduce to the familiar expression for uniform-density
Malmquist bias, $E(r|d)=de^{7\Delta^2/2}$ (e.g., Lynden-Bell
\etal\ 1988); basically, objects are more likely to
be farther away than $d$ because there is more volume
at larger distances. However, for
realistic samples this
is not always a good approximation, as the
density can vary rapidly along the line of sight. The
overall Malmquist bias arising from both the volume effect 
and density variations is known as {\em inhomogeneous
Malmquist bias,} or IHM.  In order to correct
properly for IHM, it is important 
that a realistic model of the density field be
substituted into equation~(\ref{eq:erD}). 

There is no perfect way to do this. One might use,
for example, redshift-space density $n(cz)$ as a substitute
for $n(r),$ but this would ignore the distorting
effects of peculiar velocity. Alternatively, one could 
estimate the real-space density from the number
density in inferred-distance space, $n(d).$ 
The latter approach is closely related to the
Landy-Szalay (1992) method of Malmquist-bias
correction, which we will 
implement elsewhere (Eldar, Dekel, \& Willick 1996).
However, for our present purposes the preferred technique
for the spiral samples
is to use a model of $n(r)$ derived from the \iras\
1.2 Jy redshift survey (Fisher \etal\ 1995),
with the effects of peculiar velocities
corrected for using linear theory (Yahil \etal\ 1991;
Strauss \& Willick 1995, \S~5.9). The advantage
of this approach is that \iras\ galaxies are
expected to be good tracers of the general spiral
density field. The disadvantage is that the 
reconstruction of $n(r)$ from redshift data is
necessarily model-dependent: it assumes that gravitational
instability is valid, and moreover requires that
a smoothing scale and
a value of $\beta\equiv \Omega_0^{0.6}/b,$ 
where $\Omega_0$ is the density parameter and $b$
is the linear bias factor 
be chosen for the reconstruction. 

While we recognize the objections that can be raised
to this procedure, we do not consider it 
to be a serious issue in practice. The effects
of density variations on the overall Malmquist
bias correction are typically smaller than the uniform-density
bias.
The relatively small differences
in the size of the correction
between the various possible reconstructions of $n(r)$
from \iras\ are smaller still. We
have chosen a reconstruction model in which $\beta=0.6,$
the velocity reconstruction assumes pure linear theory,
and a Gaussian smoothing scale of 300 \kms\ was used. 
The value of $\beta=0.6$ was adopted based on
a maximum likelihood comparison of a subset of
Mark III spirals with the \iras\ density and velocity
fields (Strauss \& Willick 1995, \S~8.1.3; Willick \etal\ 1997).
The smallest smoothing scale possible is optimal
for Malmquist bias correction, and
300 \kms\ is the smallest that can effectively
be used in the reconstruction. Further details
of the \iras\ reconstruction method are given
by Yahil \etal\ (1996). 

In practice, equation~(\ref{eq:erD}) is not 
especially robust for numerical calculation 
of Malmquist corrections
owing to the log-normal factor in the integrands. Instead,
we use the following,
completely equivalent, expression for $E(r|d)$ (Willick 1997):
\begin{equation}
E(r|d)\,=\,de^{\frac{7}{2}\Delta^2}\,\frac{1 + \frac{1}{\sqrt{\pi}}\infint
\delta\left(de^{4\Delta^2}
e^{\sqrt{2}\Delta x}\right)\,e^{-x^2}\,dx}
{1 + \frac{1}{\sqrt{\pi}}\infint
\delta\left(de^{3\Delta^2}
e^{\sqrt{2}\Delta x}\right)\,e^{-x^2}\,dx}\,,
\label{eq:erDint0}
\end{equation}
where $\delta(r)=n(r)/n_0 - 1$ is the fractional galaxy overdensity.
Equation~\ref{eq:erDint0} is simple to integrate because
of the strict Gaussian factor and the use of $\delta$ rather
than density itself. Furthermore, in this form one clearly
sees that the IHM correction implicitly contains the standard
homogeneous Malmquist term.
All Malmquist corrected distances
in Table 3, to be discussed in the next section,
are obtained by numerical evaluation of equation~(\ref{eq:erDint0}).

%

\subsection{The Effect of Redshift Limits on Malmquist Bias Corrections}
\def\bfv{{\bf v}}

The Malmquist bias corrections discussed above assume that sample objects
can lie at any distance along the line of sight.
This is reflected in the limits of integration---zero to infinity---in
equation~(\ref{eq:erD}).  Many current TF samples, however, 
do not have this property, because in addition to magnitude or diameter
limits, sample selection may depend on redshift as well. Restrictions
on redshift may be imposed either by observational constraints 
(e.g., \hi\ receivers
are limited in frequency range) or sample definition (e.g., observers make TF
measurements only for objects with known redshifts less than a chosen
value).  In this section we discuss a modification to the IHM correction
in the presence of a redshift limit, and comment on how such
considerations apply to the Mark III spiral samples.


The fact that a redshift limit modifies the nature of Malmquist bias
has been recognized by other workers. Freudling \etal\ (1995), for example,
modeled the effect of redshift limits using a Monte Carlo procedure, and
da Costa \etal\ (1996) used these models in constructing peculiar velocity
maps from their $I$-band TF sample. 
In contrast, we have taken an analytic approach to bias corrections.
Such an approach
has the advantage that
the assumptions
and model parameters that go into it are more evident, and their
effect on the final corrections more easily assessed, 
than in a Monte Carlo scheme. We present the outlines
of our approach below. However, for reasons desecribed
in \S~5.1.2, we have not actually made redshift limit
corrections in the Mark III database. The discussion
to follow is designed to enable users to do so
should they deem it necessary
for their particular analysis.

\subsubsection{Analytic Approach to the Redshift Limit Correction}

To modify the IHM correction for a redshift limit, one must first
model the redshift-distance relation in the vicinity of the limit. This
entails making assumptions about the peculiar velocity field,
as did the IHM correction
itself (see above). 
In the discussion
to follow we assume the redshift-distance relation near a limit is at most
a bulk departure from uniform Hubble flow.  
It is also necessary to adopt a value for the
small-scale velocity ``noise,'' $\sigma_v,$ about
the mean flow. A reasonable value is
$\sigma_v=200\ \kms$ (cf.\ Davis, Nusser, \&
Willick 1996 for further discussion).

\def\czlim{cz_{{\rm lim}}}
\def\czl{cz_\ell}
Suppose that a particular subsample is subject to a limit $cz \leq \czlim.$
Suppose furthermore that the bulk flow of galaxies near $\czlim$ (and
in the part of the sky under consideration) is given
by $\bfv_p.$ Then the first-order effect of the redshift limit is to exclude
all objects at {\em distances\/} greater than 
$\czl\equiv\czlim-\bfv_p\cdot\nhat,$ where $\nhat$ is a unit vector along the
line of sight. Note that the distance limit for a given redshift limit
is direction-dependent.  
However, the presence of
velocity noise means that 
objects whose observed radial velocities place them at the redshift limit
actually lie within a  range ($\sim \czl \pm \sigv$ )
of distances. Thus, rather than an abrupt distance limit at $\czl,$
there is a fuzzy limit, and we cannot simply replace the upper limit
of integration by $\czl.$ Instead, we multiply the integrands
in both numerator and denominator of equation~(\ref{eq:erD})
by the probability, $P(r|\czlim,\bfv_p,\nhat),$ that an object at distance
$r$ along line of sight $\nhat$ satisfies the redshift limit criterion.
This probability is given by
(Willick 1997)
\begin{equation}
P(r|\czlim,\bfv_p,\nhat) = \onehalf\left[ 1 - \erf\left(\frac{r-\czl}{\sqrt{2}\sigv}
\right)\right]\,,
\label{eq:prczl}
\end{equation}
where ``$\erf$'' is the error function. In the limit
$\sigv\rightarrow 0,$ $P(r|\czlim,\bfv_p,\nhat)\rightarrow \Theta(\czl - r).$
In other words, when $\sigv$ is very small in comparison with other relevant
scales (in this case, the TF distance error $\Delta d$)
the effect of multiplying by $P(r|\czlim,\bfv_p,\nhat)$ differs little
from replacing the upper limit of integration by $\czl.$ Similarly, it is
clear that when $\czl-r \gg \sigv,$ $P(r|\czlim,\bfv_p,\nhat)\simeq 1,$
i.e., far from the redshift limit (relative to the velocity noise) the standard
Malmquist formula is recovered. 

\subsubsection{Redshift Limit Effects in the Mark III Spiral Samples}

As noted above,
we have not taken account of redshift limit effects in computing
the IHM-corrected distances tabulated in the Mark III catalog
(\S~6.2). 
We have, in effect, taken $\czlim \rightarrow \infty$ for all of the
Mark III objects. This is not to say that redshift limit effects are 
entirely absent in the selection of
the Mark III samples. Rather, as noted in \S~2, these limits are
in most cases\footnote{The exceptions are the cluster samples, HMCL and
W91CL, as discussed below.} so ill-defined as to preclude a well-defined
correction without making explicit, \aposteriori\ cuts on the samples.
This is in fact what we do in POTENT (Hudson \etal\ 1995; Dekel \etal\ 1996):
we identify a redshift beyond which redshift-selection
effects appear to become important in each sample
(see below), and then eliminate from the analysis all galaxies at
higher redshifts. 
Equation~(\ref{eq:prczl}) then strictly applies to the reduced samples.
Since the distributed catalog includes {\em all\/} galaxies in the original
samples, however, we believe it would be misleading to adopt 
such hard redshift cuts
in computing IHM corrections for the catalog. 
The discussion above
should enable potential users of the catalog to account for redshift-limit
effects if they so choose. To allow such calculations to be made,
we include 
at the Mark III distribution sites (cf.\ \S~6.4) 
the density grid,
$n(r),$ toward each Mark III spiral. 

We now discuss the redshift selection criteria that affect the makeup
of the Mark III spiral samples. 
There is one spiral sample for which we know that no
redshift limit effects are present: the CF sample, 
which was selected strictly
on the basis of magnitude and diameter limits (Courteau 1992, 1996). 
For the remaining field samples, redshift selection effects
of a more or less pronounced character apply, as follows:

\begin{enumerate}
\item A82 exhibits an abrupt reduction in number
of objects per unit redshift at
$cz_\odot=3000\ \kms$ (Paper II, \S~6). In the POTENT
analysis, 
only A82 galaxies with $cz_\odot\leq 3000\ \kms$ are
used. The POTENT IHM correction accounts for this effect according
to the prescription outlined above
(see Paper V for further details).
There are 59 A82 galaxies at heliocentric redshifts 
$>3000\ \kms$ presented in the on-line Mark III catalog.
Users should be aware that the IHM corrections presented
for these objects are thus indicative only, as
the sample is strongly incomplete at $cz_\odot>3000\ \kms.$

\item Any redshift limits affecting MAT are
very weak. Mathewson \etal\ (1992) indicate that
their sample is confined ``in general'' to radial
velocities $< 7000\ \kms,$ but this appears to
be a consequence of the sample diameter limit (cf.\ Paper II,
\S~2.1) rather than a redshift limit {\em per se.} 
Mathewson \etal\ further indicate that in the GA region,
a number of fainter galaxies at higher redshifts were
included. Again, however, these more distant objects appear
to have been selected by relaxing the diameter limit
rather than by explicitly selecting on redshift. Thus,
to a good approximation 
the MAT Mark III sample is not redshift limited. 
However, users are
advised that this statement is probably rigorously
true only if the diameter limited nature of the
sample is respected, i.e., if small ($D_{{\rm ESO}}<1.6'$)
MAT galaxies are excluded from the analysis.

\item The W91PP sample was not subjected to a redshift limit by Willick
(1991). However, it is an \hi-selected sample 
based on the Arecibo observations
of Giovanelli \etal\ (1985,1986). The observations were implicitly
limited by the
prevailing restrictions on the Arecibo receivers at the time. 
W91PP is thus effectively
redshift-limited at \sm 12,000 \kms. This limit is applied in the 
POTENT IHM
correction for W91PP. 

\end{enumerate}

For the cluster Mark III samples, of course, the situation is
quite different. HMCL and W91CL are composed of galaxies
that were {\em expressly selected to lie with
a narrow ($\sim 1500\ \kms$) range of redshifts centered on the mean cluster
redshift.} The effect on the IHM correction of
such redshift cuts
extremely strong. As a result, the Malmquist-corrected
forward TF distances for individual HMCL and W91CL
galaxies presented in the Mark III catalog are
not applicable in a Method I analysis of these samples.
We have included them for purposes of completeness only. 
Cluster galaxies should not, in any case, be treated
individually in Method I analyses, but should be
grouped together and corrected for
selection bias, as we have done in the spiral groups
catalog (\S~6.3). 

\section{Final TF Relations and Partial Presentation of the Spirals Catalog}

In this section we present illustrative portions of the
Mark III Catalog. Because of its large size, the full
catalog will be made available electronically only,
as we describe below.
We present data for both individual spiral galaxies (the singles
catalog, \S~6.2) and groups of
spiral galaxies (the groups catalog, \S~6.3). 
In \S~6.4 we describe how to access the complete, on-line versions
of the catalog. First, however, we revisit 
our Paper II determination of inverse TF zero points,
and present a corrected, final tabulation of the TF
relations for the Mark III spiral samples.

\subsection{Corrected Inverse TF Zero Points and Final TF Relations}

We have modified slightly the
inverse TF zero points presented in Paper II,
after recognizing a problem with our earlier
approach.
In Paper II, \S~6, we applied
the same reasoning to the forward and inverse relations,
minimizing individual galaxy distance modulus
differences to 
determine final zero points. However, while this approach
ensures consistency of raw inverse TF distances {\em between\/}
samples, it does not guarantee consistency of forward and
inverse distances {\em within\/} samples. There is no need for
forward and inverse {\em individual\/} galaxy distances to agree within
a sample,
because forward and inverse distances are subject to substantially
different Malmquist bias corrections (e.g., Strauss \& Willick 1995, \S~6.6.5).
However, once corrected for selection bias,
forward and inverse {\em group\/}
distances should agree within a sample. Each is, in principle, an unbiased
measure of the distance to the group, to which no further correction
is necessary in a Method II analysis. 

However, we found in preparing the catalog that there were
systematic offsets between forward and inverse group distances
within each Mark III sample (except HMCL; see below). For example, the inverse TF
distances to the W91CL clusters
were 0.04 mag larger, in the mean, than the corresponding forward TF distances,
when the inverse zero point obtained in Paper II was used.
The origin of these differences is not entirely clear. While they
are small in an absolute sense, they are typically twice as large as
the relative zero point errors we estimated in Paper II, Table 12,
and thus significant. Their existence requires us to decide which
criterion of homogeneity we value more: agreement of individual galaxy
inverse distances between samples, or of forward and inverse
group distances within samples.

Our view is that the latter criterion is more basic, and we used
it to redetermine the inverse TF zero points for each sample except HMCL. 
Specifically, we adopted the inverse zero point
that minimized a $\chi^2$ statistic formed from
forward minus inverse TF distance moduli and errors assumed
to scale as $n^{-\onehalf},$ where $n$ was the number of
objects in the group or cluster. 
For HMCL, however, the original inverse zero point was determined in the same
way as the forward zero point---zero net cluster motion,
cf.\ Paper I, \S~3.2.2---and thus required, in principle, no further
adjustment. Application of the $\chi^2$ minimization
procedure to the HMCL forward and inverse cluster distances confirmed that
this was indeed the case.
In redetermining the inverse zero points we did {\em not\/} change
the inverse slopes or scatters from their Paper II values.
The new procedure resulted
in a small ($\simlt 0.05$ mag) changes in the inverse 
zero points. In most cases, the sense of this adjustment was
to make the inverse TF distances slightly (\sm 2\%) smaller. 
We emphasize that while the new inverse zero points have not been determined
{\em directly\/} by the overlap method, ultimately the overlap principle
governs their values: the overlap method was used to determine
forward TF zero points, and inverse distances
are now normalized by forward ones. 

The one exception to the procedure just described was
the CF sample, for which we formed no independent groups.
In this case, we simply assumed that the difference between
the forward TF zero point $A$ and the inverse TF zero point $D$
for CF was the same as for the W91CL sample. This assumption is
justified on the grounds that W91CL and CF exhibit
very similar TF relations
(Paper II, \S~4).

Having corrected the inverse TF zero points as just described,
we list the parameters of the final TF relations for the
Mark III spiral samples in Table~\ref{tab:finaltf}. Note that,
with the
exception of the modified inverse zero points, this
table is identical to Table 12 of Paper II.

\subsection{The Spiral Galaxy Singles Catalog}

In Table~\ref{tab:mat01} we present data
for 45 galaxies in the MAT sample. The format of this printed
table is the same as that of the complete electronic tables.
In the on-line version of the catalog, there is a separate
file for each sample. Each has an identical format, however,
so the portion of the MAT table presented here will provide
sufficient guidance.

The galaxies are listed in order of ascending heliocentric
redshift in Table~\ref{tab:mat01}. Column 1 lists the {\em Mark
III Catalog internal
identification number} for the object. These numbers reflect the
order in which the original authors listed their objects, typically
in order of increasing Right Ascension. For example, the first
and second entries in Table~\ref{tab:mat01} were the 1322nd and
sixth entries, respectively, in the 
data table presented by Mathewson \etal\ (1992). The number presented
in Column 1 is unique within each sample. Thus, specifying
the Mark III sample (e.g., MAT, HMCL, etc.) and the internal
identification number uniquely specify a Mark III object.
The internal identification number also facilitates cross-referencing
between the singles catalog itself, and the files of auxiliary
data for the catalog objects that are also to be found
in the electronic data base.
The remaining columns in Table~\ref{tab:mat01}
are as follows:

\noindent{Column 2: PGC Number.}

\noindent{Column 3: Name of the galaxy as listed by the original authors.
In the on-line A82 catalog, the names of the
22 Virgo Cluster galaxies whose heliocentric radial
velocities were set to 1153 \kms\ (the mean Virgo value) 
in the application of the grouping algorithm (cf.\ Paper II, 
\S~5.2.2) are followed by ``[V]''.
}

\noindent{Column 4: Group number of the galaxy. This number corresponds
to the groups listed in Table~4 (see below). For the cluster
samples (HMCL and W91CL), all objects have a group number unless
they were explicitly excluded from the TF calibration procedure
(cf.\ Paper I). The latter objects have group number $-1.$
For the field samples (W91PP, CF, MAT, A82),
objects that were placed into groups by the grouping
algorithm of Paper II have group numbers $\geq 1.$ Group
number zero signifies that the algorithm attempted to group
the object but could not because it did not have neighbors
sufficiently close in redshift space. Group number $-1$
signifies that the object was excluded \apriori\ from
the grouping procedure. For example, as explained in
Paper II, \S~2, the grouping algorithm was not applied
to objects with ESO diameters smaller than $1.6',$
with $\eta<-0.42,$ and with inclinations less than $35\degs.$
In addition, a small number of objects was excluded \apriori\
for what were judged to be unreliable axial ratios, even
if they were nominally large enough that the derived inclination
was $>35\degs.$ Although the CF sample was not grouped in Paper II,
CF objects that lie in the Perseus-Pisces region, and which
are not present in the W91PP sample, were consolidated 
with W91PP for the purpose of forming maximal groups
for later velocity analysis.
The resulting grouped sample is known as ``WCF.'' 
CF and W91PP sample group numbers correspond to the
WCF grouped sample.
}

\noindent{Column 5: Galactic longitude (degrees).
}

\noindent{Column 6: Galactic latitude (degrees).
}

\noindent{Column 7: Circular velocity parameter $\eta$ (equation~\ref{eq:defeta}).
}

\noindent{Column 8: Apparent magnitude $m$ (mag), fully corrected
for extinction and redshift (cf.\ \S~2.1).
}

\noindent{Column 9: ESO blue angular diameter (arcminutes), in the
case of the MAT sample, which is illustrated here. However,
more generally this column contains the variable upon which
sample selection was based: UGC blue diameters in the case of
CF and W91PP; UGC blue diameters or Zwicky apparent magnitudes
for HMCL North, ESO blue diameters for HMCL South; RC3 $B$ magnitudes
for A82.
}

\noindent{Column 10: Total correction from raw to corrected apparent
magnitude $\Delta m$ (mag), 
as described in \S~2.1. 
The quantity $\Delta m$ is
$>0$ when the corrected apparent magnitude is smaller
(brighter) than the raw magnitude (the usual case). 
The case $\Delta m < 0$ can occur because we correct
to a fiducial axial ratio (rather than face-on) for
W91 and CF (cf.\ \S~2.1.3), and also because
we derived a negative internal extinction coefficient
for A82 (cf.\ Paper II, \S~5.2.7).
}

\noindent{Column 11: $B$-band Galactic extinction (mag; \S~2.1.2).
}

\noindent{Column 12: Logarithm of the (major-to-minor) axial ratio ${\cal R}$
(\S~2.1.3).
}

\noindent{Column 13: For all samples except A82, this column
lists the Burstein Numerical Morphological Type (BNMT). This index is
a numerical encoding of the RC3 morphological type, developed
by one of us (DB). A detailed description of 
the BNMT is given in Appendix B. For the A82 sample, the BNMT
was not available, and the RC2 numerical morphological type
is listed instead.
}

\noindent{Columns 14--16: Three measures of the TF distance to the
object, all given in \kms.
Column 14 gives the raw forward TF distance.
Column 15 gives the forward TF distance corrected for
IHM, as described in \S~5. Column 16 gives the raw inverse
TF distance.
For reasons described in Strauss \&
Willick 1995 (\S~6.5.5), the inverse distances have a more
complicated Malmquist bias correction, which we consider
elsewhere (Eldar, Dekel, \& Willick 1996).
}

\noindent{Columns 17--19: Radial velocities in \kms, as measured in the
heliocentric ($v_\odot$), Local Group ($v_{{\rm LG}}$), 
and Microwave Background ($v_{{\rm CMB}}$) frames
of reference, respectively. The heliocentric velocities are
those measured by the original authors except for a
few cases, discussed in \S~3, where the overlap
comparison revealed a deviant value, in which case
the deviant values are replaced by the valid ones.
We transform from heliocentric
to Local Group velocities according to the transformation
of Yahil \etal\ (1979). CMB-frame velocities are obtained
using the motion of the sun with respect to the CMB 
determined by the COBE dipole anisotropy 
(Kogut \etal\ 1993).
}

\noindent{Column 20: The expected distance in \kms, $d_{{\rm IRAS}},$
derived from the same \iras\ reconstruction as was
used in the Malmquist correction procedure (\S~5).
This distance was computed as the expectation
value of true distance, given the observed radial velocity
and the \iras-predicted peculiar velocity and density fields. A small-scale
velocity dispersion of $150\ \kms$ was assumed in the
calculation. See Strauss \& Willick (1995, \S~8.1.3) for
further explanation. 
}

\noindent{Column 21: The local galaxy overdensity $\delta,$ defined as
$(n_g-n_0)/n_0,$ where $n_g$ is the local number
density and $n_0$ is the mean number density, again obtained
from the \iras\ reconstruction. The \iras\ 
density was evaluated at the IHM-corrected forward TF distance 
when $v_{{\rm LG}}<750\ \kms,$ and at the \iras-expected
distance otherwise.
}

\noindent{Column 22: The forward TF residual, $\delta m_{{\rm TF}},$ in mag.
This residual was computed with respect to the TF fits to the
groups formed by the grouping algorithm of Paper II (W91PP, CF, MAT, A82)
or the cluster TF fits of Paper I (HMCL, W91CL). As noted above,
W91PP and nonoverlapping CF galaxies in Perseus-Pisces
were grouped together for the purposes of this compilation. 
When an object either was not included in the grouping
algorithm or cluster fits (e.g., MAT objects with $D_{{\rm ESO}}<1.6'$
or CF objects away from PP), or was not placed in a group by
the algorithm because of a lack of redshift-space neighbors,
there is no TF residual for the object; the value in column 22
is then given as $-9.999.$
}

\subsection{The Spiral Groups Catalog}

In Table~\ref{tab:matg}, we present representative data
from the grouped portion of the Mark III catalog spirals. For
consistency with Table~\ref{tab:mat01}, we present here
groups formed from the MAT sample. 
In the on-line version of the catalog, grouped
data are also presented for HMCL, W91CL, A82, and W91PP plus
Perseus-Pisces CF galaxies (WCF).
HMCL and W91CL galaxies were grouped \apriori\ based on assumed cluster
membership (Paper I). 
The field samples (MAT, A82, WCF) were grouped by
the grouping algorithm (cf.\ Paper II, \S~2.2.2). 

The groups in Table~\ref{tab:matg}, and in the on-line
catalog, are listed approximately in order of ascending heliocentric 
redshift\footnote{The grouping algorithm initially sorted objects
on heliocentric redshift, and the field
sample singles files are thus listed precisely in this order. However,
in the process of grouping there is some inevitable shuffling
back and forth; as a result, the groups themselves are not listed 
{\em exactly\/}
in order of their mean heliocentric redshifts.}
for the field samples. For the cluster samples the order reflects
the convention originally adopted for HMCL (cf.\ Paper I, Table~1;
the HMPP clusters are listed at the end of the HMCL list).
Column 1 is the group number, which uniquely identifies a
group within each sample. This number corresponds to
that listed in column 4 of Table~\ref{tab:mat01};
it is thus straightforward to identify the individual
members of the group by cross-referencing the two tables.
Column 2 lists the number of
galaxies in the group, $N_g.$ Columns 3 and 4 list the mean Galactic
longitude ($l$) and latitude ($b$) of the group members. 
Column 5 lists the mean velocity width parameter, $\bar{\eta},$
of the members of the group. Column 6 is the rms scatter, 
in mag, of the group members about the TF relation. (The
TF relation fitted to the group is the universal TF relation
for the sample, not a fit just to the
members of the group.) Column 7 lists the forward TF distance
to the group, $d_{{\rm TF}},$ in \kms. This distance is fully
corrected for selection bias. Because the groups are formed
using redshift-space criteria, it is selection rather than
Malmquist bias which pertains (cf.\ Strauss \& Willick 1995, \S~6.4).
Column 8 lists the inverse TF distance to the group,
again corrected for selection bias (although in the inverse
case the correction is extremely small; cf.\ the relevant
discussions in Papers I and II). As noted above, the inverse
TF zero points were chosen so that the forward and inverse
TF group distances agree in the mean, although significant
differences are occasionally seen in individual cases.
Column 9 lists the distance modulus error $\delta\mu_{{\rm TF}}$ (mag)
associated with the TF distance; it is computed simply as
$\sigma_{{\rm TF}}/\sqrt{N_g},$ where $\sigma_{{\rm TF}}$ is
the magnitude scatter of the TF relation for the
sample in question (e.g., 0.43 mag MAT).
Columns 10, 11, and 12 list, respectively, group heliocentric, LG,
and CMB frame radial velocities
in \kms. The heliocentric group radial velocities are computed
as the mean heliocentric radial velocity of group members if $N_g=2,$ and
as the median radial velocity for groups with three or more members;
the LG and CMB frame group radial velocities are then obtained by
transforming the heliocentric group radial velocity as described
in \S~6.1.

\subsection{The Electronic Catalog}
 
The Mark III Catalog has been made available electronically at three
separate sites. The first is NASA's Astronomical Data Center, which
may be accessed either using a Web browser or by anonymous FTP at
{\tt adc.gsfc.nasa.gov.} The second site is a Web page maintained by
Willick, at the URL {\tt http://astro.stanford.edu/MarkIII.} The third
site is an anonymous FTP resource maintained by Burstein at
{\tt samuri.la.asu.edu.} The contents of these three archives are
very similar, although slight differences of organization
exist. At each site extensive documentation in the form of README files
is available.

The files are given in ASCII format and
are organized into five main directories: (i) individual spiral galaxy
data, (ii) spiral group and cluster data, (iii) spiral ``overlap galaxy'' data,  
(iv) spiral ancillary data, and (v) elliptical galaxy
data. The first directory contains data files named {\tt mark3\_mat\_s, mark3\_w91cl\_s,}
and so forth. These files correspond to Table~4 of this paper, and are
described by a single README file called {\tt RMk3\_ind\_dist.}
The second directory contains data files called {\tt mark3\_mat\_g, mark3\_w91cl\_g,} and
so forth. They correspond to the information in Table~5 of this paper, and
are described by a single README file called {\tt RMk3\_gp\_dist.} The
third directory contains just one data file, {\tt mark3\_match,}
which is nearly identical in content
to Table~2 of this paper (the electronic version does not
contain Galactic coordinates) and is described by the README
file {\tt RMk3\_match.} The fourth directory contains a set of data files not
described in this paper. 
There are three separate data files for each Mark III spiral sample (there is
no file of ancillary data for the elliptical galaxies), named {\tt matfileX.lst,}
where {\tt X=1,2,3,} described
by README files called {\tt RMk3\_mat,} and so forth.
These files contain data that are not crucial to peculiar velocity analyses,
but which might be useful for related studies, including
apparent magnitudes
and angular diameters from a variety of galaxy catalogs and cross-referencing
information between catalogs. In addition, these
files list the original photometric and velocity width data
as reported by the original Mark III sample authors. Finally, in
the fifth directory one may find the elliptical galaxy data. These
data are presented in exactly the same manner as they were in
the Mark II catalog distributed in 1989 by Burstein: there are two
data files, {\tt egalfile1.lst\/} and {\tt egalfile2cor.lst,} and
a single README file called {\tt RMk3\_egal.} These files differ
from the Mark II distribution only by the small multiplicative
correction to the \dnsigma\ distances, 
as described in \S~7 below.

On the Web page maintained by Willick, two additional 
types of data are available. First, as mentioned in
\S~5.1, the (normalized) \iras\ galaxy number densities
$n(r),$ with values given at quadratically spaced positions
along the line of sight
toward each Mark III spiral, are provided.  A short FORTRAN program
to read the binary files containing this information is also made available.
Second, twenty realizations of simulated Mark III catalogs, generated
as described by Kolatt \etal\ (1996), may be found, along with documentation
describing their use.

\section{Matching the Elliptical and Spiral Distances}
\def\vx{\pmb{$x$}}
\def\pmb#1{\setbox0=\hbox{#1}%
 \kern-.025em\copy0\kern-\wd0
 \kern.05em\copy0\kern-\wd0
 \kern-.025em\raise.0433em\box0}
\def\vx{\pmb{$x$}}

The sample of elliptical and S0 galaxies with \dnsigma\ distances
is added almost as is from the Mark II data set compiled by D. Burstein
(based on Lynden-Bell \etal\ 1988; Faber \etal\ 1989; Lucey \& Carter 1988;
Dressler \& Faber 1990). It includes 544 galaxies in 249 objects 
(single galaxies, groups and clusters).
However, we first rescaled the Mark II \dnsigma\ distances
in order to match the elliptical and spiral distances,
as we now describe.

The original \dnsigma\ zero point
was determined independently, and is therefore not necessarily consistent
with the global TF zero point determined in Paper I.
We thus allow for a multiplicative degree of freedom in the \dnsigma\
distances, $d \rightarrow (1+\epsilon) d$, corresponding
to a free Hubble-like monopole component in the peculiar velocities,
$u \rightarrow u - \epsilon r$. The value of $\epsilon$ is determined
subject to the assumption that both the ellipticals and the spirals are
unbiased tracers of the same underlying velocity field
(for a discussion of the validity of this assumption see Kolatt \& Dekel
1994, hereafter KD).
We found in three different ways that the best value is $\epsilon = -0.035
\pm 0.01$, and have corrected the \dnsigma\ distances accordingly before adding
them to the catalog.

One method of matching is described in detail in KD.
The same large-scale POTENT smoothing was applied separately to the TF and the
\dnsigma\ data, yielding two independent radial peculiar velocity fields,
$u_s(\vx)$ and $u_e(\vx)$, and their corresponding errors,
$\sigma_s(\vx)$ and $\sigma_e(\vx)$, at common grid points $\vx$.
The POTENT smoothing mimics a
spherical Gaussian window of radius $1200\ \kms$ with minimum biases
due to the sparse and nonuniform sampling of noisy radial velocities
(Dekel, Bertschinger \& Faber 1990;
Dekel 1994; Dekel \etal\ 1996).  The two velocity fields were then
compared at grid points near which the sampling by both types of
galaxies is ``adequate", which we define as having at least five neighboring galaxies
of the same type within a sphere of radius $1500\ \kms$.
The sampling by the ellipticals limits
the comparison to a volume of an effective radius $\sim 4000\ \kms$.
The two fields were matched by minimizing the statistic
\begin{equation}
D=\sum \left.\left[ {(u_e-u_s)^2\over\sigma_e^2}
             +{(u_e-u_s)^2\over\sigma_s^2} \right]  \right/
  \sum \left[ {(u_e+u_s)^2\over\sigma_e^2}
             +{(u_e+u_s)^2\over\sigma_s^2} \right] \, ,
\label{eq:Des}
\end{equation}
where the sum is over the adequate grid points.
The comparison at grid points together with the inverse weighting by the
errors is a compromise between the desired equal-volume weighting and
the optimal treatment of noise.

This analysis was applied in KD to a preliminary version of the
Mark III data, and it was re-done recently using the final version of the
catalog, with a very little change in the result.
The best fit values range between $\epsilon = -0.05$ and $-0.02$,
depending on the exact volume of comparison. The
correction is statistically significant despite the fact that it is small.
Based on the distribution of $D$ in Monte Carlo simulations, the probability
that the elliptical and spiral velocity fields
are both noisy versions of the same underlying field is more than 10\%
after an $\epsilon=-0.035$ correction, while it was less than 2\%
before the correction.

In an alternative analysis, the radial peculiar velocity of each elliptical
galaxy (or group) was compared with the average of the radial velocities 
of the neighboring spirals inside a top-hat sphere of radius $500\ \kms$.
In this analysis the effective smoothing is on much smaller scales,
thus reducing the biases within the effective window to a level where they
can be practically ignored. However, this comparison is not volume weighted.
The best fit is found again to be $\epsilon=-0.035$ with similar errors.

In a third analysis, the inferred distances of the ``same"
elliptical and spiral clusters were compared.
It turns out that there are severe difficulties in trying to identify
matching clusters. The spiral ``clusters" in many cases extend
over several Mpc and only a handful of them can be confidently
identified with elliptical counterparts.
Even when the identification is quite certain, the different types
of galaxies may show different mean velocities because they
tend to sample different components of the cluster. We ended up with
6 clusters in common, 
and with a best fit of
$\epsilon\simeq -0.03$, fully consistent with the other tests.

\section{Further Consideration of the TF Residuals}

Several assumptions underlie most statistical analyses of TF-type
data. The most frequently adopted 
are the following:
\begin{enumerate}
\item TF residuals are Gaussian; 
\item TF residuals are independent of
velocity width; 
\item  TF residuals are uncorrelated with morphological type.
\end{enumerate}
In this section we subject these assumptions to simple
but stringent tests using three samples: MAT, A82, and WCF.
We will
conclude that the first and third of the above assumptions
are consistent with our data. The second assumption clearly fails
for the MAT sample, but to a much lesser degree (if at all) for
the other two; we discuss possible reasons for this and
suggest how velocity analyses might account for this effect.
\begin{figure}[th]
\centerline{\epsfxsize=4.0 in \epsfbox{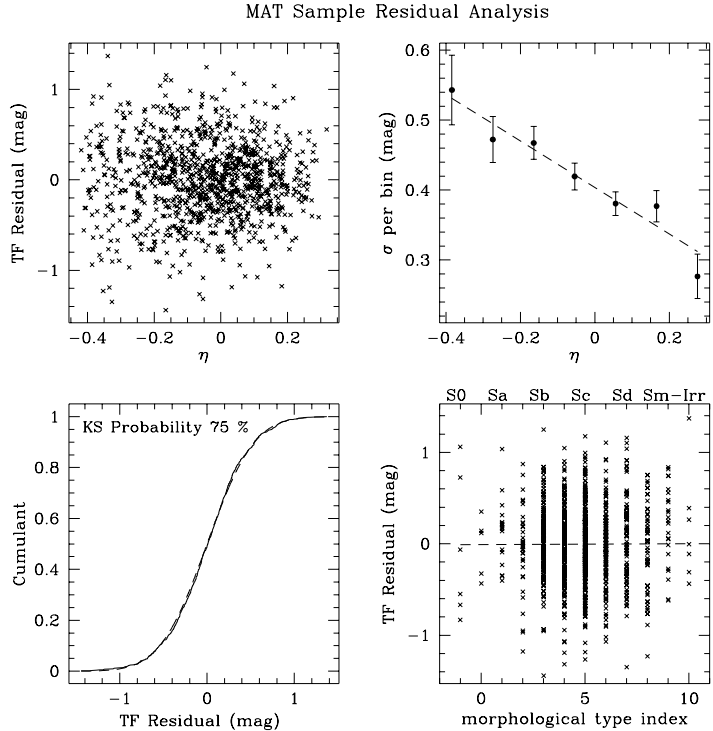}}
\caption{Upper left panel: MAT TF residuals are plotted
versus the velocity width parameter $\eta.$ Upper right panel:
rms values of the TF residuals, computed within bins of
width $\Delta\eta=0.11,$ are plotted against $\eta.$ The
dashed line shows the best-fit linear relation between
the TF $\sigma$ and $\eta.$ Lower left
panel: the cumulative distribution (normalized to
unity) of the TF residuals
(solid line) and the corresponding distribution for
a Gaussian with the same rms dispersion (dashed line)
are plotted. The Kolmogorov-Smirnov probability
that the distributions are the same is indicated.
Lower right panel: TF residuals are plotted versus
RC3 morphological type index; the dashed
line shows the best fit linear relation between the
mean residual and the type index. See text for details.}
\label{fig:matresid}
\end{figure}

\begin{figure}[th]
\centerline{\epsfxsize=4.0 in \epsfbox{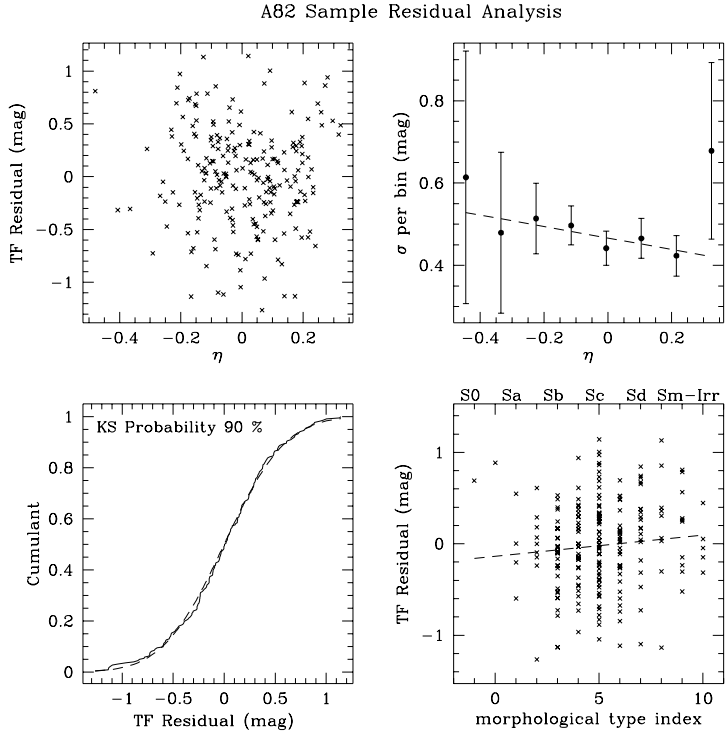}}
\caption{Same as Figure 2, except that
the residual analysis is now done for the
A82 sample.}
\label{fig:a82resid}
\end{figure}

\begin{figure}[th]
\centerline{\epsfxsize=4.0 in \epsfbox{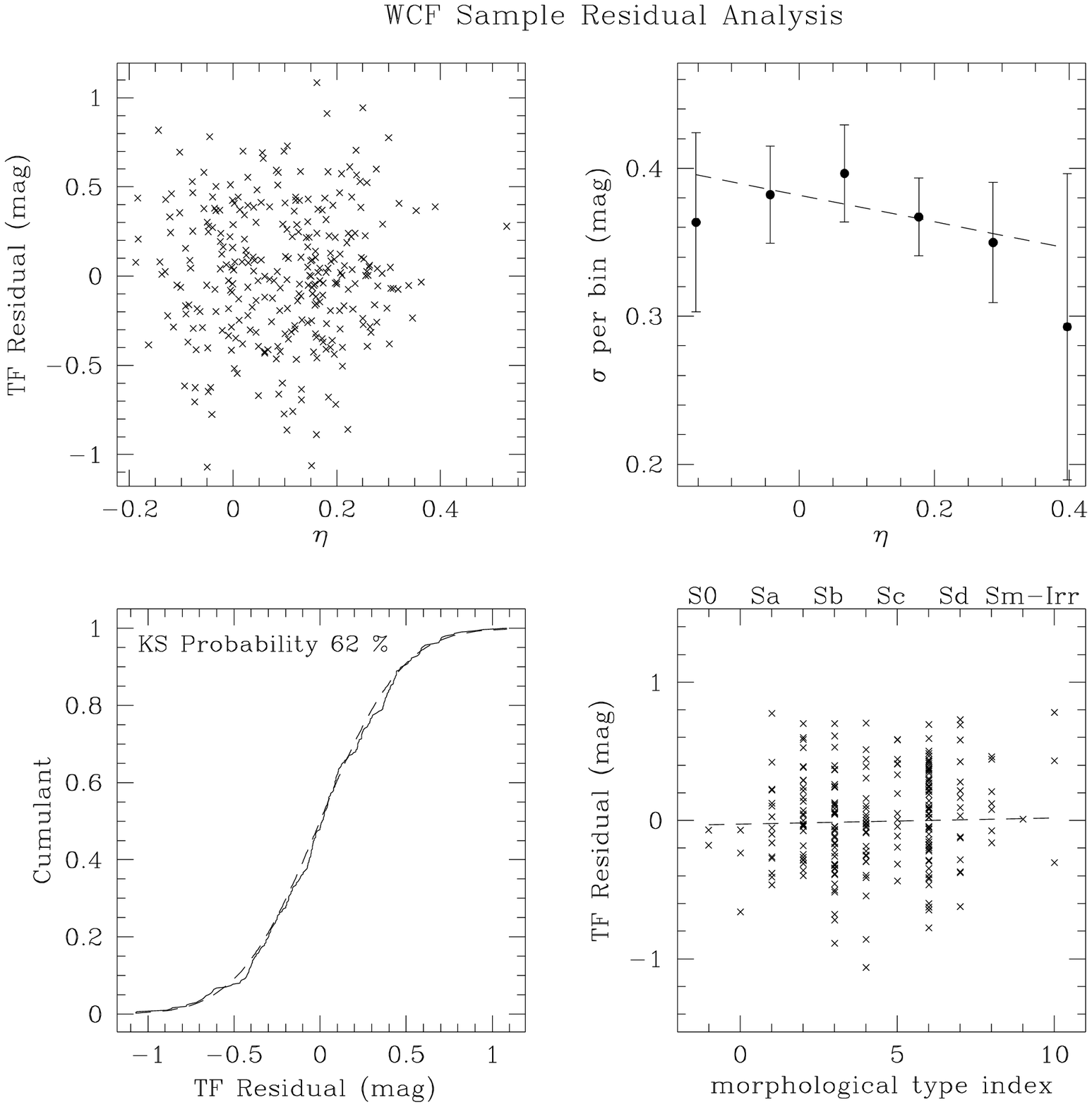}}
\caption{Same as Figure 2, except that
the residual analysis is now done for the
combined W91PP+CF sample.}
\label{fig:wppresid}
\end{figure}

For each of the three samples, we use the TF residuals
computed by the grouping algorithm (cf.\ Paper II, \S~2.2.2)
and tabulated in Table~3 (or the corresponding electronic file). 
These residuals
are plotted versus $\eta$ in the upper left panels
of Figures~\ref{fig:matresid}, \ref{fig:a82resid},
and~\ref{fig:wppresid} for MAT, A82, and WCF
respectively. 
The advantage of the
grouping algorithm residuals (as compared with the
HMCL and W91CL cluster fit residuals) is that the assignment
of objects to groups was done objectively.\footnote{Recall
from Paper II (\S~2.2.2) that the grouping algorithm used an
``input'' TF scatter to reject objects from group membership.
Thus, extreme ($\simgt 3.3\sigma$)
outliers are, in effect, already excluded from the
present analysis. Were this not done, the TF residuals
would not be strictly Gaussian. Our view is that
Gaussianity of the residuals is sufficiently desirable
as to warrant the exclusion of a small (\sm 1--2\%) number
of sample objects.}
Our test for Gaussianity utilizes the Kolmogorov-Smirnov (KS)
statistic. 
The KS test measures the probability that 
the cumulative distribution of the residuals
is drawn from a Gaussian distribution with the same dispersion as
the rms value of the residuals themselves. The results of the KS tests are
plotted in the lower left panels of 
Figures~\ref{fig:matresid}--\ref{fig:wppresid}.
The cumulative distributions of the residuals
are plotted as solid lines; the cumulative distributions
of the corresponding Gaussians (with dispersions 0.43 mag
for MAT, 0.47 mag for A82, and 0.38 mag for WCF)
are plotted as dashed lines. It is visually apparent
that the two curves are in good agreement in each case.
The value of the KS probability is indicated in each panel as
a percentage.
For all three samples, the KS probability is large ($\simgt 60\%$),
whereas a large deviation from Gaussianity would be signified by
a small value ($\simlt 10\%$). From these results we 
conclude that the 
assumption that TF residuals are Gaussian
is justified.

Next, we consider whether the TF scatter is constant
with velocity width (or, equivalently, with luminosity).
For each of the three samples, we have
computed the rms value of the TF residuals within
bins of width $\Delta\eta=0.11.$
The results are plotted in the upper right panels
of Figures~\ref{fig:matresid}--\ref{fig:wppresid}. 
In the case of the MAT sample, a clear trend
is seen: $\sigma$ decreases with increasing $\eta,$ i.e., 
the TF scatter is smaller for bright galaxies than it
is for faint galaxies. The dashed line represents the
best-fit straight line to the binned rms values.
For the MAT sample, this straight line is given by
$\sigma(\eta) = 0.404\,(.008)-0.33\,(.05)\eta,$
where \onesigma\ errors 
are indicated in parentheses. The nonzero slope
that characterizes the trend is highly significant.
For the A82 and WCF samples, a qualitatively similar
trend is seen.
However, in each of these two cases, the
fitted slopes differ from zero at only about the
$1.5\,\sigma$ significance level. Specifically, for
A82 the relation is $\sigma(\eta) = 0.466\,(.013)-0.14\,(.09)\eta.$
For WCF it is $\sigma(\eta) = 0.382\,(.010)-0.09\,(.06)\eta.$
Thus, the decrease in TF scatter with increasing velocity
width is not unambiguously detected in the A82 and WCF samples.

Given the strong trend seen in the MAT sample,
one must reinterpret the KS test for the Gaussianity
of the MAT TF residuals. Because $\sigma$ is not
constant with $\eta,$ the residuals cannot obey
a strictly Gaussian distribution. However, their
overall distribution irrespective of $\eta$ can
still be Gaussian if both the TF residuals
at any given $\eta,$ and the
distribution of $\eta$-values, are Gaussian. It is
difficult to test the latter assumption because of
selection effects. We have, however, performed KS tests
on the residuals within each $\Delta\eta=0.11$ bin
shown in Figure~\ref{fig:matresid}. We find that within
nearly all bins, the KS probability that the residuals
obey a Gaussian distribution with scatter $\sigma(\eta)$
is $\simgt 50\%.$ The one exception is the bin centered
at $\eta=-0.16,$ in which the KS probability is 2.6\%.
Inspection of the upper left panel of Figure~\ref{fig:matresid}
reveals the reason for this result---a scarcity of
residuals in the range $\sim -0.1$--$0$ mag, 
and an excess of residuals of $\sim +0.2$ mag.
The reason for this deviant bin is unknown.\footnote{It is
worth noting, however, that if one analyzes a sequence
of $N$ Gaussian distributions for Gaussianity, the
probability of finding one that appears {\em non-Gaussian}
at a given significance level is proportional to $N.$
It is thus not necessarily significant that one of
the seven bins tested exhibits non-Gaussian behavior.}
Excepting
this unaccounted-for behavior, our results indicate
that it is valid to view the MAT TF scatter as Gaussian
at any given $\eta,$ but as a linear function $\sigma(\eta)$
as described above. This ``local'' Gaussianity ensures that
the statistical techniques we have applied in this
series of papers remain valid. 

We have not explicitly carried out a test for Gaussianity
using inverse TF residuals. However, such an exercise is
unnecessary, because the forward and inverse residual
distributions are in fact closely related. 
In Appendix C, we apply the laws of probability theory
to derive the distribution of inverse TF residuals 
from those
of the forward relation. We show that local Gaussianity
of the forward residuals implies local Gaussianity of
the inverse TF residuals as well---as long as the change
in scatter with velocity width is gradual, 
and the luminosity function is wider than
the TF scatter. These conditions are shown to hold
quantitatively for actual TF samples. Statistical
techniques that assume local Gaussianity are therefore
valid for inverse, as for forward, TF analyses.
We also discuss in Appendix
C the factors which cause the inverse TF relation 
to differ from the mathematical inverse of the
forward TF relation---i.e., which result in $D\neq A,$ and $e\neq b^{-1}$---properties
of the observed TF relations that were previously unexplained.
 
The rather different scatter versus $\eta$ behavior
evidenced by MAT, as compared with A82 and WCF,
represents an ambiguous result.
If TF scatter were inherently a strongly decreasing function
of $\eta,$ we would expect to see the trend in all samples.
But the $\sigma(\eta)$ versus $\eta$ slopes for A82
and WCF differ from that of the MAT sample
at the \sm 2- and 3$\,\sigma$ levels respectively.
This raises the possibility that the
trend seen in the MAT sample is an artifact of that data set. 
Alternatively,
it could be that only the MAT sample is large
enough, and in particular rich enough in low-linewidth
objects, that an actual trend can be clearly detected.
It is worth noting that, if velocity width errors $\delta(\Delta v)$
are roughly independent of the width itself, then $\eta$ errors
$\delta\eta \propto \delta(\Delta v)/\Delta v$ increase
with decreasing velocity width. If so, the part of the
TF scatter due to width errors ($\sim b\delta\eta,$ where
$b$ is the forward TF slope), and
thus the TF scatter itself, must similarly increase with
decreasing $\eta.$ Thus, at some level, the trend seen
in the MAT data is bound to occur. Whether an additional
effect of real physical significance (related, e.g., to
galaxy formation physics) is present is difficult to
say. For the present, the most prudent approach is
to examine the effect 
of allowing the TF scatter to vary with velocity width,
in any given peculiar velocity analysis. However, the
allowed variation should be constrained
by the results obtained above, e.g., 
the $\sigma$ versus $\eta$ relation should be
taken as linear with the slopes calculated above.
We will adopt this approach in future papers
(Dekel \etal\ 1996; Willick \etal\ 1997). In
the on-line catalog, however, we have assumed
fixed TF scatter independent of velocity width
in computing the Malmquist corrections (\S~6.2). 
The values of the TF scatter adopted are those
given in Table~\ref{tab:finaltf}.

Finally, in the lower right panels of Figures 2--4
we plot TF residuals versus RC3 
morphological type index. 
This measure of morphology runs from $\simlt 0$ for
very early-type spirals (S0, S0a, Sa) to $\simlt 10$
for the late-type spirals. 
Dwarf galaxies, multiple galaxies, and galaxies
with highly uncertain morphology 
are not shown and
are not considered in the analysis to follow.
It is apparent in each case that the
TF residuals do not correlate, or correlate
at most very weakly, 
with galaxy morphology. 
To quantify this impression, we have carried out
linear fits of the mean residual within each 
bin to the numerical index. The dashed lines 
show the results of these fits.
For MAT, the slope of the fitted line is $0.001\pm 0.009;$
for A82, it is $0.023\pm 0.014;$ and
for WCF, it is $0.005\pm 0.005.$
Thus, for MAT and WCF we may
confidently reject the presence of significant correlation
between the TF residuals and morphological type. For
A82, a weak trend 
may be present. It is possible that the trend is real
for A82, which is based on aperture magnitudes,
but is eliminated through the use of CCD total magnitudes.
Given its marginal significance, a more conservative
assumption is that the trend is negligible for A82,
as it clearly is for MAT and WCF.
Thus, the Mark III data do not support
the notion that galaxies of different morphological
types obey distinct TF relations. 
This conclusion is unaffected if we
restrict the analysis to the relatively large
objects, $D\geq 2.5\arcmin,$ whose
morphologies are least uncertain.

\section{Transforming the Samples to a Common System}

An important principle underlying the calibration procedure
of Papers I and II was that each individual sample
had a distinct TF relation. This was understood 
as a consequence of the distinct character
of each data set: $I$- versus $r$- versus $H$-band photometry,
different measures of velocity width, etc. Indeed
the TF parameters were found to differ markedly among the samples
( Table~\ref{tab:finaltf}).
However, for some purposes it is inconvenient
to have more than one TF relation
involved in a velocity analysis. For example, the
approach to velocity
field reconstruction developed by Nusser \& Davis (1995), 
and applied to the Mark III Catalog by Davis, Nusser, \&
Willick (1996), is greatly simplified if
the entire sample obeys a single TF relation. 
In order for a catalog consisting of disparate samples
to have this property, the TF observables (apparent
magnitude and velocity width)
for at least some of the spiral samples must be
suitably transformed.\footnote{We make no effort
to incorporate the ellipticals into this scheme.}
In this section we derive such transformations for
the Mark III spiral samples.

As we did in finalizing TF distances (cf.\ Paper II, \S~6),
we take HMCL as a template. That is, all apparent
magnitudes and velocity widths will be transformed 
to an ``HMCL-equivalent'' system, henceforth the {\em common system.}
The basic idea is the following: we assume that for each sample ($S,$ say)
other than HMCL, the velocity widths $\eta_S$ and apparent magnitudes
$m_S$ can be transformed to their HMCL-equivalent values according
to relations of the form
\begin{equation}
\eta_{\rm common} = a_{0,S}+a_{1,S}\eta_S+a_{2,S}\eta_S^2
\label{eq:etaequiv}
\end{equation}
and
\begin{equation}
m_{\rm common} = m_S + b_{0,S} + b_{1,S}(m_S-5\log r)\,.
\label{eq:mequiv}
\end{equation}
The possibility that the coefficient $a_1$ differs from
unity arises because velocity width systems differ
as to the precise quantity they measure.
The quadratic term in equation~(\ref{eq:etaequiv}) 
was found to differ from zero only in the case of
the CF sample, as we discuss further below.
A non-zero value
of the coefficient $b_1$ allows for a
luminosity dependence of galaxy color in the case that
sample $S$ is not based on $I$-band photometry.
As we show below, $b_1$ differs significantly from zero
only for the $H$-band A82 sample.

Proceeding in analogy with Paper II, \S~6, we obtain the
coefficients in equations~(\ref{eq:etaequiv}) and~(\ref{eq:mequiv})
through a prioritized overlap comparison. The 
objects used in this comparison are those listed in Table~2.
We first consider objects common to
HMCL and W91CL, and determine the transformation coefficients
for the latter sample by
fitting the HMCL data (widths and magnitudes separately) to the W91CL
data by least squares. 
The fits are initially carried out assuming that all coefficients
in equations~\ref{eq:etaequiv} and~\ref{eq:mequiv} are
potentially significant. However, when the initial fits fail to
yield values of certain coefficients that differ
significantly from zero, those coefficients are assumed
to be identically zero and the fits are redone without
them. That is, we assume that the data sets are as alike
as they can possibly be, and only allow nonzero coefficients
when these are forced upon us by the data. 

Once the transformation is determined for W91CL, all
W91CL magnitudes and velocity widths are transformed to 
their common system values. We then compare W91PP objects
with their counterparts in both HMCL and W91CL, thus determining
the transformation of W91PP to the common system.
The CF sample is then compared with HMCL and the transformed
W91CL and W91PP and its transformation determined; MAT is
then compared with HMCL and the transformed W91CL, W91PP, and CF,
and finally A82 is compared with HMCL and the transformed
W91CL, W91PP, CF, and MAT. Each comparison yields the
coefficients in equations~(\ref{eq:etaequiv}) and~(\ref{eq:mequiv})
that allow a transformation to a common system. There
is one exception to the hierarchy just described, however.
Previous comparisons have shown full consistency between
the W91CL, W91PP, and CF photometry (Willick 1991; Courteau 1992;
Courteau 1996). 
Thus, in determining the magnitude transformation
law, W91CL, W91PP, and CF are grouped together and compared
with HMCL. For the velocity width transformation, however,
these samples are treated separately for reasons discussed
in Paper II.

Table~\ref{tab:commonsys} 
summarizes the results of these overlap comparisons.
Note that the ``transformation'' coefficients for HMCL are
trivial, as HMCL defines the common system.
Several aspects of Table~\ref{tab:commonsys} warrant further comment.

\begin{enumerate}
\item The W91CL sample ought, by construction, to be on the
HMCL $\eta$-system. The raw velocity widths used by W91CL and
those used by Han \& Mould (1992) for their northern clusters
(which overlap completely with W91CL; see Paper I, Table 1)
are the same, namely, those tabulated by Bothun \etal\ (1985).
Any systematic difference between the HMCL and W91CL $\eta$'s
would therefore imply a systematic difference in the derived inclination
corrections, and, thus, in the measured axial ratios. The
fact that 
$a_0=0$ and $a_1=1$ for W91CL is thus indicative of a consistency
between the W91CL and HMCL axial ratio assignments. 
W91PP and CF aimed at full consistency with the
HMCL $\eta$-system. Their 
non-zero values of $a_0$ indicate a marginally
significant discrepancy.

\item We present both a linear and a
quadratic $\eta$-transformation for
CF (fourth and fifth lines of Table~\ref{tab:commonsys}). 
The quadratic fit results in a small
reduction in scatter,
and largely eliminates a trend seen in residuals
from the linear fit. 
It is not surprising that the CF velocity
widths, which are optically measured (Courteau 1992),
are not as simply related to the \hi\ 21 cm widths of
HMCL and W91 as the latter are with one another.
The quadratic transformation
for the CF widths is used in the common-system analysis
of Davis, Nusser, \& Willick (1996). However, in the
TF calibration of the CF sample presented in Paper II, 
and in the distributed Mark III catalog,
no transformation of the CF widths (nor those
of any other sample) is made.

\item The coefficient $a_1$ for the MAT sample differs
significantly from unity. This effect is a consequence
of the very different definition of \hi\ velocity width
used by Mathewson \etal\ (1992) from that
of the Aaronson group (see, e.g.,
Bothun \etal\ 1985). The nature of the transformation
is such that the MAT $\eta$-value is markedly smaller
for a faint galaxy than the common system $\eta$ for
the same object; however, for the brightest galaxies
($\eta\simgt 0.3$) the MAT $\eta$ differs little
from the common system value. This effect
also explains why the MAT TF
relation (cf.\ Table 12 of Paper II) is so
much flatter than the HMCL TF relation; the ratio
of the MAT to the HMCL TF slope is $0.86$, very close
to the value of $a_1$ for the MAT sample in Table~\ref{tab:commonsys}. 

\item The origin of the large photometric zero point offset between the MAT 
and common system (i.e., 
HMCL) apparent magnitudes (the
coefficient $b_0$ in line 6 of Table~\ref{tab:commonsys})
is not well understood. 
Both MAT and HMCL carried out Kron-Cousins $I$-band photometry.
However, the offset is unmistakable; the coefficient $b_0$
differs from zero at the $6\,\sigma$ significance level.
It is thus essential to transform the MAT magnitudes
to bring them to the common system. We note that
this magnitude transformation, in combination
with the width transformation discussed above, 
fully accounts for the difference between the
MAT and HMCL TF relations (Table~\ref{tab:finaltf}). 

\item  The A82 magnitude transformation (row 7 of Table~\ref{tab:commonsys})
has a coefficient $b_1$ that differs significantly from zero.
This term is a consequence of a strong luminosity dependence
of the $(I-H)$ colors of galaxies. Note that the size of this
coefficient is very nearly what is expected from the
difference between the $I$-band ($b_I = 7.87 \pm 0.16$) 
and $H$-band ($b_H = 10.29 \pm 0.22$) TF slopes (cf.\ Paper II, Table 12),
i.e., $(1-b_1)^{-1}\times b_I \simeq  b_H.$ 
In deriving the coefficient $b_1$,
it was necessary to estimate the distances $r$ to A82 galaxies
in carrying out the overlap fit (see equation~\ref{eq:mequiv}).
This was done by taking $r=cz_\odot$ for $cz_\odot> 100\ \kms,$
and setting $r=100\ \kms$ otherwise. This procedure, while
imperfect, suffices for the purposes of the fit.

\item The A82 velocity widths are slightly offset, 
by $0.016$ in $\eta,$ from the common system widths.
The origin of this offset is unknown, as both sets of widths
stem from the work of the Aaronson group in the 1980s,
and both measure width at 20\% of the peak of the \hi\ profile.
Nonetheless, it is a clearly detected ($\sim 5\,\sigma$) effect.
Because the widths of Ursa Major galaxies in W91CL were
derived principally from the A82 sample, we have augmented
W91CL Ursa Major galaxy
width parameters by $0.016$ in the Mark III singles catalog.
This is necessary for W91CL Ursa Major galaxies to be
mutually consistent with the remainder of the W91CL sample.
The width increase
has the effect of increasing W91CL Ursa Major distances
by $7.73\times 0.016 = 0.123$ mag over their original values.
This distance increment largely accounts for the discrepancy
originally seen in the A82 versus
W91CL overlap comparison (cf.\ Paper II, \S~6).

\end{enumerate}


\section{Summary and Further Discussion}

We have presented a number of technical details
concerning the construction of the Mark III Catalog of
Galaxy Peculiar Velocities. Our procedures for converting
raw apparent 
magnitudes and velocity widths into corrected
values suitable for application of the TF relation were described.
We presented the full list of overlap galaxies that
allow us to bring together disparate spiral samples for
peculiar velocity studies and reviewed the means
by which elliptical galaxy \dnsigma\ data are zeropointed
consistently with the spirals. We discussed our technique for 
computing
inhomogeneous Malmquist bias corrections for spirals and
indicated how such corrections can break down in
the vicinity of redshift limits.
Inverse TF zero points 
were rederived based on the requirement that
forward and inverse TF group distances agree
within each sample. The final TF relations for
the Mark III spiral samples are given in Table~\ref{tab:finaltf}.
We presented abbreviated versions
of the Mark III catalog, 
and provided potential
users with a guide to 
accessing the electronic
catalog in \S~6.4.

A simple analysis of the properties of TF residuals was
presented. We confirmed one of the widely-made assumptions
about the TF relation, namely, that it exhibits Gaussian
residuals. In the case of the MAT sample, however, we
found that while the residuals are Gaussian at any given
velocity width, their rms value $\sigma(\eta)$ is an
approximately linearly decreasing function of $\eta,$
i.e., the TF scatter decreases with increasing luminosity.
This has been suggested elsewhere (e.g., Federspiel, Sandage, \&
Tammann 1994; Freudling \etal\ 1995) but never conclusively
demonstrated. The WCF and A82 samples exhibited qualitatively
similar but much weaker trends with marginal statistical significance.
We found no evidence for a meaningful dependence of the TF
relation on morphological type across the entire range (Sa--Sd)
of spirals well represented in these samples.  

Our chief concern in constructing the Mark III Catalog has been
ensuring the uniformity of the data and the proper calibration of
the individual sample TF relations. Toward this end, we have modified
the observable data presented by the original authors, because we
have applied our own, uniform set of corrections to the raw data.
More importantly, we have substantially changed the derived
TF distances as compared with the original authors because
we have recalibrated the TF relations characterizing each data set.
Thus, velocity analyses based on the Mark III catalog
will differ significantly from, and should be considerably more
reliable than, comparable analyses based on the original data.

We cannot be certain, however, that the final catalog is
entirely free of systematic errors. A crucial link in our
chain of reasoning is that the HMCL sample data are uniform
between its Northern and Southern sky components. Any unaccounted
for discrepancy between the photometric or \hi\ properties
of HMCL South as compared with HMCL North could vitiate that basic
assumption.
Another variable we cannot fully control is possible
redshift-dependencies of the basic data in any given sample.
For these reasons, it is essential that observational
checks on the uniformity of the catalog be carried out
in the future. Three of the present authors (JAW, SC, and MAS),
along with D.\ Schlegel (Durham) and M.\ Postman (STScI),
are carrying out a full-sky TF survey of galaxies in
the redshift range 4500--7000 \kms, one of whose aims is
to test for and correct possible systematic errors in 
Mark III. Comparison with other TF surveys (e.g., Giovanelli
\etal\ 1996) will also be important.
We will attempt to disseminate 
results from these studies in a timely fashion.

\acknowledgments
The authors would like to thank Jeremy Mould and Ming-Sheng
Han for their cooperation in our efforts to assemble and
tabulate their cluster data set, and Don Mathewson for
making his extensive TF sample available on computer tape.
We also acknowledge the contributions of Amos Yahil in
developing various methods of velocity and density
field reconstruction from the \iras\ redshift survey.
The work presented here was supported in part by 
NSF Grant AST90-16930 to DB and
by the US-Israel Binational Science Foundation.
This research has made use of the NASA/IPAC Extragalactic
Database (NED) which is operated by the Jet Propulsion Laboratory, California
Institute of Technology, under contract with the National Aeronautics and Space
Administration.

\appendix
\vfill\eject

\section{The Cosmological Correction for Tully-Fisher Magnitudes}

\def\zref{z_0}
\def\zrat{5\log\left(\frac{1+z}{1+\zref}\right)}
\def\nrat{2.5\log\left[\frac{\inf0int N(\frac{\lambda}{1+\zref})\,S(\lambda)\,d\lambda}{\inf0int N(\frac{\lambda}{1+z})\,S(\lambda)\,d\lambda}\right]}
As discussed in \S~2.1.4, the standard K-correction
is not appropriate for apparent magnitudes used in
peculiar velocity studies.\footnote{The discussion
to follow applies to the CCD samples (HMCL, W91, CF, MAT) only.
The $H$-band A82 sample requires a different K-correction, as
discussed in \S~2.1.4).} Whereas
the standard correction is applied
to apparent magnitudes so that they
yield luminosity-distances, 
our correction must instead lead to an estimate of
the ``cosmological redshift'' $z_c$ and the associated
distance in \kms\ units, $r=cz_c.$ It is this distance
that, when compared with the observed redshift (in
velocity units) $cz,$ yields a peculiar velocity estimate.

To obtain the desired correction,
we begin with the monochromatic energy flux observed from a 
galaxy at redshift $z$ (e.g., Peebles 1971):
\begin{equation}
f(\lambda)=\frac{L\left(\frac{\lambda}{1+z}\right)}{4\pi a_0^2\,x^2(1+z)^3}
\label{eq:flambda}
\end{equation}
Here $L(\lambda)$ is the spectral energy distribution of the
galaxy in its rest frame, $a_0$ the present day scale factor
of the universe, and $x$ the comoving coordinate distance
of the galaxy, which is related to its cosmological redshift
$z_c$ by the equation 
\begin{equation}
a_0 x = \frac{c}{H_0}\left[\frac{q_0 z_c + (q_0 - 1)(-1 + \sqrt{2q_0 z_c + 1})}{q_0^2 (1+z_c)}\right] \equiv \frac{c}{H_0}Z_q(z_c) 
\label{eq:defr}
\end{equation}
(\eg, Weinberg 1972). In equation~(\ref{eq:defr}), $q_0$
is the deceleration parameter, 
and the convenient notation $Z_q(z)$ 
has been borrowed from Schneider, Gunn,
\& Hoessel (1983). It is important to note that
while the cosmological redshift $z_c$ determines
the value of the coordinate distance $x,$ the
observed redshift $z$ in equation~(\ref{eq:flambda})
incorporates not only the expansion of the universe
but also the peculiar motions of the observer and the source;
to sufficient accuracy it is simply the heliocentric
redshift of the galaxy.

The magnitudes we are concerned with here are total
magnitudes measured with a CCD, and hence depend
not on the energy flux $f(\lambda),$ but
instead on 
the photon number flux $n(\lambda).$ 
The energy of a single photon of wavelength $\lambda$ is
$e(\lambda) = hc/\lambda,$
where $h$ is Planck's constant, and therefore
$n(\lambda)=f(\lambda)/e(\lambda)\propto \lambda f(\lambda).$
Similarly, the photon luminosity in the galaxy rest frame
is related to its energy luminosity by 
$N(\lambda)\propto \lambda L(\lambda).$
Combining these relations with equation~(\ref{eq:flambda})
one obtains the following proportionality:
\begin{equation}
n(\lambda) \propto \frac{N(\frac{\lambda}{1+z})}{x^2(1+z)^2}\,.
\label{eq:defnlam}
\end{equation}

The CCD apparent magnitude is related to the number flux
$n(\lambda)$, integrated over the transmission curve, $S(\lambda)$, of the
bandpass in question:
\[ m=C-2.5\log\inf0int n(\lambda)S(\lambda)\,d\lambda \]
\begin{equation}
= C' -2.5\log\inf0int N\!\left(\frac{\lambda}{1+z}\right)S(\lambda)\,d\lambda + 5\log\,(1+z)+5\log Z_q(z_c).
\label{defm}
\end{equation}
In our system of units (cf.\ Paper I, \S~2.1), 
absolute magnitude is defined as the apparent magnitude
at a distance of $1\ \kms,$ corresponding to a redshift
of $\sim 3\times 10^{-6}$. For the time being, let us
denote this minuscule redshift $\zref$;
we will drop this quantity at the end, 
but it is convenient to maintain it in the derivation
that follows. With this convention, it follows that
the absolute magnitude of an object with photon luminosity
$N(\lambda)$ is given by
\begin{equation}
M = C' -2.5\log\inf0int N\!\left(\frac{\lambda}{1+\zref}\right)S(\lambda)\,d\lambda + 5\log\,(1+\zref)+5\log Z_q(\zref)\,.
\label{eq:defM}
\end{equation}

Using equations~\ref{defm} and~\ref{eq:defM} we see that the
galaxy distance modulus, not yet corrected for cosmological
effects, is given by
\begin{equation}
 m-M =2.5\log\left[\frac{\inf0int N(\frac{\lambda}{1+\zref})S(\lambda)\,d\lambda}{\inf0int N(\frac{\lambda}{1+z})S(\lambda)\,d\lambda}\right] + 5\log\left(\frac{1+z}{1+\zref}\right) + 5\log\frac{Z_q(z_c)}{Z_q(\zref)}\,.
\label{eq:defmuuc}
\end{equation}
In order to obtain the desired cosmological correction,
we must now ``unpack'' $z_c$ from the complicated 
expression $Z_q(z_c)$ in equation~(\ref{eq:defmuuc}).
Expanding equation~(\ref{eq:defr}) through first order in $z_c,$ we find
\begin{equation}
5\log\frac{Z_q(z_c)}{Z_q(\zref)} \simeq  5\log\frac{z_c}{\zref} -1.086 (1+q_0)(z_c - \zref).
\label{aplzq}
\end{equation}
Similarly, we expand the term containing the observed
redshift in equation~(\ref{eq:defmuuc}): 
\begin{equation}
\zrat \simeq 2\times 1.086 (z - \zref). 
\label{eq:defzapr}
\end{equation}
An approximation for the term involving the photon luminosity may
be derived by noting that, to first order in $z$,
$N\!\left(\frac{\lambda}{1+z}\right) \simeq N(\lambda - \lambda z) \simeq N(\lambda) - \lambda z N'(\lambda).$
Using this expansion one then finds, after some algebra,
\begin{equation}
2.5\log\left[\frac{\inf0int N(\frac{\lambda}{1+\zref})S(\lambda)\,d\lambda
}{\inf0int N(\frac{\lambda}{1+z})S(\lambda)\,d\lambda}\right]
\simeq 1.086\,\frac{\inf0int \lambda N'(\lambda)S(\lambda)\,d\lambda}{\inf0int N(\lambda)S(\lambda)\,d\lambda}\,(z - \zref)\,.
\label{eq:nterm}
\end{equation}
We may simplify further by noting that, as galaxy spectra are quite
smooth in the red,
it is reasonable to approximate $N(\lambda)$ as 
a power law for wavelengths within the $R$ and $I$ 
bandpasses.
If we thus write 
$N(\lambda) \simeq N(\lambda_{{\rm eff}})\left(\frac{\lambda}{\lambda_{{\rm eff}}}\right)^\epsilon,$
then 
$\lambda N'(\lambda) = \epsilon N(\lambda).$
Substituting this into equation~(\ref{eq:nterm}) gives us the simplified approximation
\begin{equation}
\nrat \simeq 1.086\,\epsilon\,(z - \zref)\,.
\label{eq:ntf}
\end{equation}

Using the approximations~(\ref{aplzq}), (\ref{eq:defzapr}), 
and~(\ref{eq:ntf}) in equation~(\ref{eq:defmuuc}), we
may rewrite the distance modulus as 
\begin{equation}
m - M = \ktf(z,z_c;\zref) + 5\log\frac{z_c}{\zref}\,,
\label{defkprime} 
\end{equation}
where
\begin{equation}
\ktf(z,z_c; \zref) = 1.086\,[(\epsilon + 2)(z - \zref) - (1+q_0)(z_c - \zref)]\,.
\label{eq:defKcorr}
\end{equation}
Equation~\ref{defkprime} is correct to first order in $z$ and $z_c$,
and therefore adequate for work at redshifts $\simlt\ 10,000\ \kms.$
In the logarithmic term on the right hand
side of equation~(\ref{defkprime}), we multiply $z_c$ and $\zref$
by $c$. By definition, $c\zref=1$, while $cz_c=r$, the distance
in units of \kms. In the remainder of the expression,
the tiny quantity $\zref$ may be dropped,
as it is
several orders of magnitude smaller than either $z$ or $z_c$.
Finally, then, we have
\begin{equation}
[m-\ktf(z,z_c)]-M = 5\log r\,,
\label{truedistmod}
\end{equation}
where
\begin{equation}
\ktf(z,z_c) = 1.086\,[(\epsilon + 2)z - (1+q_0)z_c]\,.
\label{defKcorr}
\end{equation}
Equation~(\ref{truedistmod}) shows that 
$K(z,z_c)$ is the proper cosmological correction for
peculiar velocity work involving the TF relation. The
practical application of this correction is described
in the main text (\S~3.1.4).

\section{Burstein Numerical Morphological Types}

The idea for developing a numerical code for the morphological types
of galaxies originated in the Second Reference Catalog of Bright Galaxies
(de Vaucouleurs, de Vaucouleurs \& Corwin 1976).  That first numerical
system simply assigned a running number from -5 to 10 to Hubble types,
with E galaxies being denoted as -5 and Irr galaxies denoted as 10.

However, once catalogs were transferred from paper to electronic means,
a more detailed numerical classification system became desirable for
several reasons.  First, the whole reason to go to a numerical scheme
is to permit easy indexing within computer programs.  Hence, the fact that
the absence of a morphological type in the catalog (a blank space) is
numerically read as a zero meant that it was desirable to assign a unique
morphological number to non-standard cases.  Second, with the placing 
of the UGC into a computer data file, and later the ESO catalog, it 
further became desirable to extend the numerical classifications to 
include objects such as multiple galaxies, dwarf galaxies, etc., for
easy computer analysis of these catalogs.  

Third, and perhaps most importantly for the UGC and ESO computer versions,
numerous differences exist in the alphanumeric characters assigned to a given
galaxy class. For example, in the ESO catalog, the subclass E/S0 alone is
written in 8 different ways, each way containing between 30 and 416 entries in
the catalog (e.g.,  E~-~S0, e~-~S0, E-~S0, E/S0, E-S0, E-S0:, S0~-~E, E~-S0). 
Each different alphanumeric rendition of the same type code is, of course,
read as different entries by a computer.  In all, the ESO catalog has
500 different alphanumeric sets of characters for the less than 40 actual
morphological classifications.  In the case of the UGC, the number is
181 separate alphanumeric sets.  The easiest way to ensure uniform 
handling of morphological types was to assign a number to each type.

As such, when Burstein first began to work with computer galaxy files
in 1977, it became desirable to define a numerical morphological code
that could both uniquely identify the different subclasses of galaxies
in the UGC, and could in principle be expanded for future catalogs if and
when more detailed information is available on galaxies.  
Hence, what we call
the Burstein Numerical Morphological Type
was developed.

The principle behind this code is to define a three-digit number.  The full
three digits gives maximal information about the object (e.g., SBa, SBa/b,
SABa, etc.).  The first two significant digits gives more general information
(e.g., E, S0, S0/a Sa, Sa/b, etc.), while the first significant digit (or
absence of it) generally separates large classes (E+S0; Spirals; Irregulars;
Dwarfs; Compacts; Multiples, etc.).  Because accessing the first significant
digit and the second two significant digits is produced by a simple integer 
division by 10 in standard programming, this code is hierarchical.  Moreover,
the existing computer catalogs of the UGC and ESO contain numerous
typographical errors. 
Assigning a unique hierarchical morphological code to each galaxy is 
necessary if accurate assessments of galaxy types in each catalog are to be
done.

The correspondence between the BNMT and the better-known RC3 morphological
type indices is presented in Table~\ref{tab:morph}.

\section{A Note Concerning Forward versus Inverse TF Residuals}

\def\nsige{\frac{1}{\sqrt{2\pi}\,\sigma(\eta)}}
\def\expmMe{\exp\!\left[-\frac{(M-M(\eta))^2}{2\sigma(\eta)^2} \right]}
In \S~8, we showed that forward TF residuals are well-approximated as
``locally'' Gaussian, i.e., Gaussian at a given value of $\eta,$ 
with rms dispersion $\sigma(\eta).$ The function $\sigma(\eta)$
was found to be a linearly decreasing function of $\eta,$ with
a slope that was significantly nonzero only for the MAT sample.

We could carry out a similar study of inverse TF residuals. However,
this is unnecessary because the forward and TF residuals are closely
related. Given the distribution of forward residuals, that
of inverse residuals follows from analytic considerations, as described 
below. While the general expressions are complicated,
we will show that under a set of assumptions reasonably
supported by the data, local Gaussianity of forward TF
residuals implies local Gaussianity of inverse TF residuals as well.

The local Gaussianity of forward TF residuals may be expressed
mathematically as
\begin{equation}
P(M|\eta) = \nsige \expmMe\,,
\label{eq:pmeta}
\end{equation}
where $M(\eta)=A-b\eta$ is the TF relation.   
We may now ask, given equation~\ref{eq:pmeta}, what is the distribution
of velocity width parameters given $M$---i.e., of inverse TF resiudals?
Using the standard rules of probability distributions we
obtain
\begin{equation}
P(\eta|M) = \frac{P(M,\eta)}{P(M)} = \frac{\phi(\eta) P(M|\eta)}{\infint\phi(\eta) P(M|\eta)}\,,
\label{eq:petaM}
\end{equation}
where $\phi(\eta)$ is the underlying distribution of velocity width parameters.

Let us write the linear scatter-width relation as
\begin{equation}
\sigma(\eta) = \sigma_0 - g\eta\,,
\label{eq:sig-eta}
\end{equation}
where $g$ was found in \S~8 to be $0.33\pm 0.05$ for MAT,
$0.14\pm 0.09$ for A82, and $0.09\pm 0.06$ for WCF.
The exponent in equation~\ref{eq:pmeta} may be expressed in terms of
\begin{equation}
\frac{M-M(\eta)}{\sigma(\eta)} = \frac{\eta+b^{-1}(M-A)}{b^{-1}(\sigma_0-g\eta)}
= \frac{\eta-\eta_0(M)}{b^{-1}\sigma(\eta_0)[1-\frac{g}{\sigma(\eta_0)}(\eta-\eta_0(M))]}\,,
\label{eq:exp1}
\end{equation}
where we have defined $\eta_0(M)\equiv -b^{-1}(M-A),$ and $\sigma(\eta_0)=\sigma_0-g\eta_0(M).$
Note that $\eta_0(M)$ is the mathematical inverse of the forward TF relation; it
is close to but not exactly equal to the inverse TF relation, as we show below.

If the term
in square brackets in the denominator of equation~\ref{eq:exp1} differed
significantly from unity, it would induce a strong
departure from Gaussianity when inserted into equation~\ref{eq:petaM}.
However, the factor $g/\sigma(\eta_0)$ is $\simlt 0.7$ for the MAT sample,
and considerably smaller for the other TF samples (cf.\ \S~8). Moreover,
the quantity $\eta-\eta_0(M)$ is restricted to lie in the range $\sim \pm 0.05,$
the inverse TF scatter. Thus, the correction represented by this term
is typically only a few percent. Given the limited accuracy with which we
can distinguish Gaussian from non-Gaussian residuals, the term is
unimportant, and we drop it in what follows.

\def\exeta0{\exp\!\left[-\frac{(\eta-\eta_0(M))^2}{2\sigma_\eta(M)^2} \right]}
If we now substitute equation~\ref{eq:exp1} into equation~\ref{eq:petaM} 
we obtain
\begin{equation}
P(\eta|M) =
\frac{\phi(\eta)\left[1-\frac{g}{\sigma(\eta_0)}(\eta-\eta_0(M))\right]^{-1}\exeta0}
{\infint \phi(\eta)\left[1-\frac{g}{\sigma(\eta_0)}(\eta-\eta_0(M))\right]^{-1}\exeta0}\,,
\label{eq:exp2}
\end{equation}
where we have reexpressed as above the $\sigma(\eta)$'s that appear outside
the exponential factors, and defined 
\begin{equation}
\sigma_\eta(M)\equiv b^{-1}\sigma(\eta_0)=b^{-1}\left[\sigma_0+gb^{-1}(M-A)\right]\,.
\label{eq:defsigeta}
\end{equation}  
This last quantity is the inverse TF scatter. It has a luminosity
dependence that corresponds to the velocity width dependence of the
forward TF scatter. Now, in view of the small effective range of
the exponential factor, $|\eta-\eta_0(M)|\simlt \sigma_\eta(M)\simeq 0.05,$
we can expand the factors that appear outside the exponential to first order:
\begin{equation}
\phi(\eta)\left[1-\frac{g}{\sigma(\eta_0)}(\eta-\eta_0(M))\right]^{-1}\simeq
\,\phi(\eta_0)\left[1 + \left(\frac{\phi^\prime}{\phi} + 
\frac{g}{\sigma(\eta_0)}\right)(\eta-\eta_0(M))\right]\,.
\label{eq:taylor}
\end{equation}
In equation~\ref{eq:taylor}, both $\phi^\prime=d\phi/d\eta$ and
$\phi$ itself are understood to be evaluated at $\eta_0(M),$ and
as such are functions of $M.$
As noted above, the term representing the luminosity dependence
of scatter is small, and higher order terms in its expansion
can safely be neglected. The other
term, $\phi^\prime/\phi,$ is the reciprocal of the effective range of
$\eta$ values, which is $\sim 0.2,$ while the inverse TF scatter is
$\sim 0.05.$ Thus, $\frac{\phi^\prime}{\phi}(\eta-\eta_0(M))$ is relatively small
($\simlt$ a few tenths). Higher order terms in the expansion of $\phi(\eta)$
will be $\simlt 10\%$ and may be neglected.

With these simplifications, we may 
substitute equation~\ref{eq:taylor} into equation~\ref{eq:exp2}
to obtain an approximation to the distribution of $\eta$ given $M.$ This leads
to an expression that contains a linear term multiplying a Gaussian. However,
to the same order of approximation used in arriving at equation~\ref{eq:taylor},
the linear term may be reexpressed as an exponential one. Its exponent
may then be combined with that of the Gaussian, and the square
completed. When this is carried out, one arrives, finally, at the following result:
\begin{equation}
P(\eta|M) = \frac{1}{\sqrt{2\pi}\sigma_\eta(M)} \exp\!\left[-\frac{\left[\eta-(\eta_0(M) + \Delta\eta_0(M))\right]^2}{2\sigma_\eta(M)^2} \right]\,,
\label{eq:petam-final}
\end{equation}
where 
\begin{equation}
\Delta\eta_0(M)\equiv\left[\frac{\phi^\prime}{\phi}+\frac{g}{\sigma(\eta_0)}\right]\sigma_\eta(M)^2\,.
\label{eq:defDeta}
\end{equation}

Equation~\ref{eq:petam-final} shows that, given our assumptions, inverse TF
residuals possess a locally Gaussian distribution. The expectation value
of $\eta$ given $M$---i.e.,
the inverse TF relation---is 
given by
\begin{equation}
\eta^0(M) = \eta_0(M)+\Delta\eta_0(M) = -b^{-1}(M-A)+\left[\frac{\phi^\prime}{\phi}+\frac{g}{\sigma(\eta_0)}\right]\sigma_\eta(M)^2\,.
\label{eq:invTF}
\end{equation}
Its scatter is $\sigma_\eta(M).$ Note that, because the $\eta$-distribution
function, its derivative, and $\sigma(\eta_0)$ are all functions of
absolute magnitude $M,$ not only is the inverse
TF zero point shifted from the ``naive'' expectation $b^{-1}A,$ but the slope
is shifted from $b^{-1}$ as well. The size of the shift depends mainly on the
logarithmic derivative of $\phi(\eta).$ For arbitrary $\phi(\eta),$
the shift is luminosity-dependent, and consequently
produces a nonlinear inverse TF relation even if the forward relation is
linear. Only in the case that $\phi(\eta)$ is Gaussian will the shift be
luminosity-independent and thus preserve the linearity of the inverse TF
relation. The fact that we cannot detect meaningful deviations from
linearity in the inverse TF relation suggests that, for TF samples at least,
$\phi(\eta)$ is well-approximated by a Gaussian distribution.
The luminosity-dependent scatter factor $g$ also will have a slight
effect on the slope, although this will be quite small. Note
that even if the forward scatter were independent
of luminosity ($g=0$), the inverse TF relation will not be the mathematical
inverse of the forward. A more detailed discussion of these issues
was given by Willick (1991, Appendix C).

In summary, we have considered the question of the distribution
of inverse TF residuals given that forward TF residuals
are locally Gaussian (\S~8). In
this Appendix we have shown that, if we make the
reasonable assumptions that (1) the change in TF scatter
with velocity width is slow, in the sense $g\sigma_\eta/\sigma_0\ll 1,$
and (2) the $\eta$-distribution function $\phi$ is wide in comparison with
$\sigma_\eta,$ inverse TF residuals are locally Gaussian as well.
The luminosity-dependence of the inverse TF scatter is straightforwardly
related to the velocity-width dependence of the forward 
TF scatter (equation~\ref{eq:defsigeta}).
Moreover, the inverse TF relation is shifted, relative to the mathematical
inverse of the forward TF relation, by an amount that depends on
$\phi(\eta),$ $g/\sigma_0,$ and the TF scatter (equation~\ref{eq:defDeta}).
The larger
the TF scatter, all other things being equal, the more the inverse TF
relation will differ from the inverse of the forward. 

\vfill\eject

\vfill\eject

\begin{table}[th]
\centerline{\begin{tabular}{l | c c r | l }
\multicolumn{5}{c}{Principal Characteristics of the Mark III Spiral Samples} \\
\hline\hline
Sample& Photometric Method & Spectroscopic Method & Number & Notes \\ \hline
HMCL      & CCD $I$-Band & \hi\ Profile Widths & 428 & a \\
W91CL     & CCD $r$-Band & \hi\ Profile Widths & 156 & b \\
W91PP     & CCD $r$-Band & \hi\ Profile Widths & 326 & c \\
CF        & CCD $r$-band & Optical Rotation Curves & 321 & d \\
MAT       & CCD $I$-band & \hi\ + Optical & 1355 & e \\
A82       & Photoelectric $H$-band & \hi\ Profile Widths & 359 & f \\ \hline
\end{tabular}}
\caption{Notes: (a) The Han-Mould Cluster sample. The
original papers describing these data are: Mould \etal\ (1991,1993);
Han (1992); Han \& Mould (1992).
The electronic catalog includes the HMPP (Han-Mould Perseus-Pisces) subset
of HMCL, which was not used in the global TF calibration; cf.\ Paper I.
(b) Willick (1991) Cluster sample. (c) Willick (1991) Perseus-Pisces
field sample (cf.\ also Willick 1990).
(d) Courteau-Faber field sample;
Courteau (1992, 1996; Courteau \etal\ 1993).
(e) Mathewson, Ford, \& Buchorn (1992) field sample.
(f) Aaronson \etal\ (1982) field sample.
A recalibration of the original A82 photometry was carried out
by Tormen \& Burstein (1995) and is adopted for the catalog.}
\label{tab:summary}
\end{table}

\begin{table}[th]
\caption{This table is contained in the file {\bf 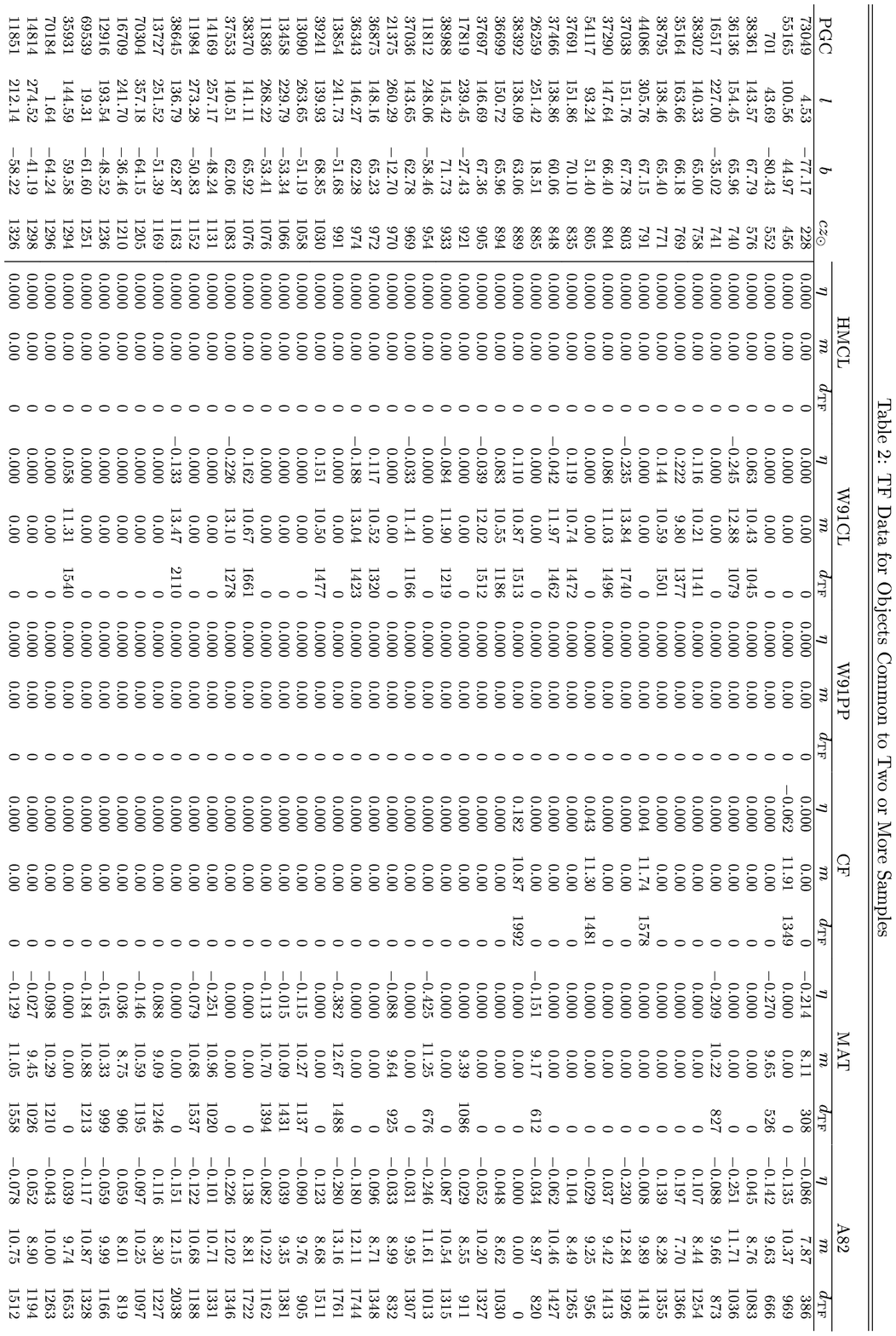.}}
\label{tab:match}
\end{table}

\begin{table}[th]
\centerline{\begin{tabular}{l | c r c c c c c}
\multicolumn{8}{c}{Final TF Relations for Mark III Samples} \\ \hline\hline
 & \multicolumn{3}{c}{forward} & & \multicolumn{3}{c}{inverse} \\ \cline{2-4} \cline{6-8}
Sample & $A\,(\pm)$ &\multicolumn{1}{c}{$b\,(\pm)$}& $\sigma$ (mag)& & $D\,(\pm)
$ & $e\,(\pm)$ & $\sigma_\eta$ \\ \hline
HM    &$-5.48\,(0.03)$ & $7.87\,(0.16)$& 0.40 & &$-5.58\,(0.03)$&$0.1177\,(0.002
5)$& 0.048 \\
W91CL &$-4.18\,(0.02)$ & $7.73\,(0.21)$& 0.38 & &$-4.23\,(0.02)$&$0.1190\,(0.003
2)$& 0.047 \\
W91PP &$-4.28\,(0.02)$ & $7.12\,(0.18)$& 0.38 & &$-4.32\,(0.02)$&$0.1244\,(0.003
1)$& 0.049 \\
CF    &$-4.22\,(0.02)$ & $7.73\,(0.21)$& 0.38 & &$-4.27\,(0.02)$&$0.1190\,(0.003
2)$& 0.047 \\
MAT   &$-5.79\,(0.03)$ & $6.80\,(0.08)$& 0.43 & &$-5.96\,(0.03)$&$0.1328\,(0.001
6)$& 0.059 \\
A82   &$-5.95\,(0.04)$ &$10.29\,(0.22)$& 0.47 & &$-5.98\,(0.04)$&$0.0893\,(0.001
8)$& 0.043 \\ \hline
\end{tabular}}
\caption{Parameters of the fully calibrated
TF relations for the Mark III spiral samples.}
\label{tab:finaltf}
\end{table}

\begin{table}[th]
\caption{This table is contained in the file {\bf 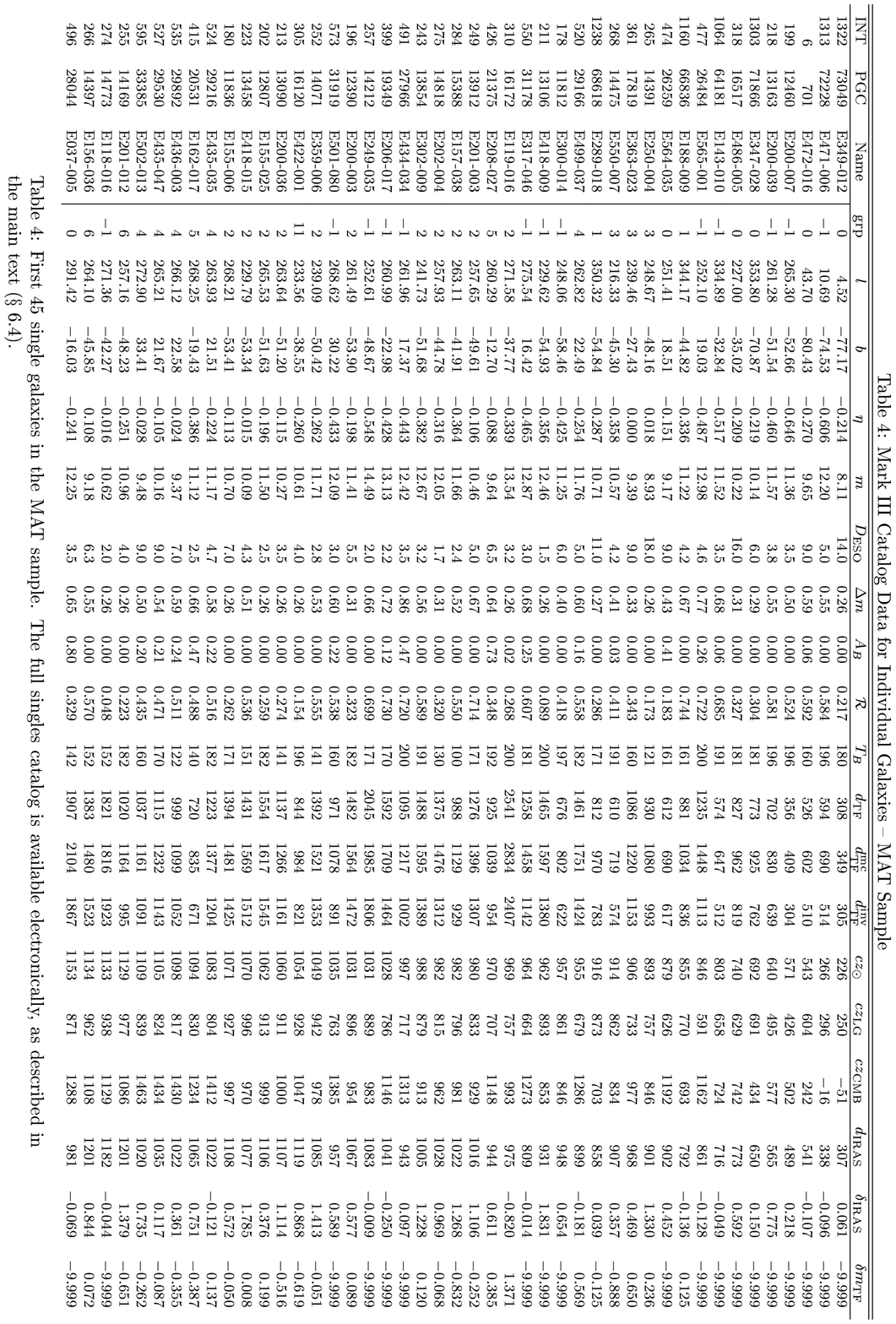.}}
\label{tab:mat01}
\end{table}

\begin{table}[th]
\caption{This table is contained in the file {\bf 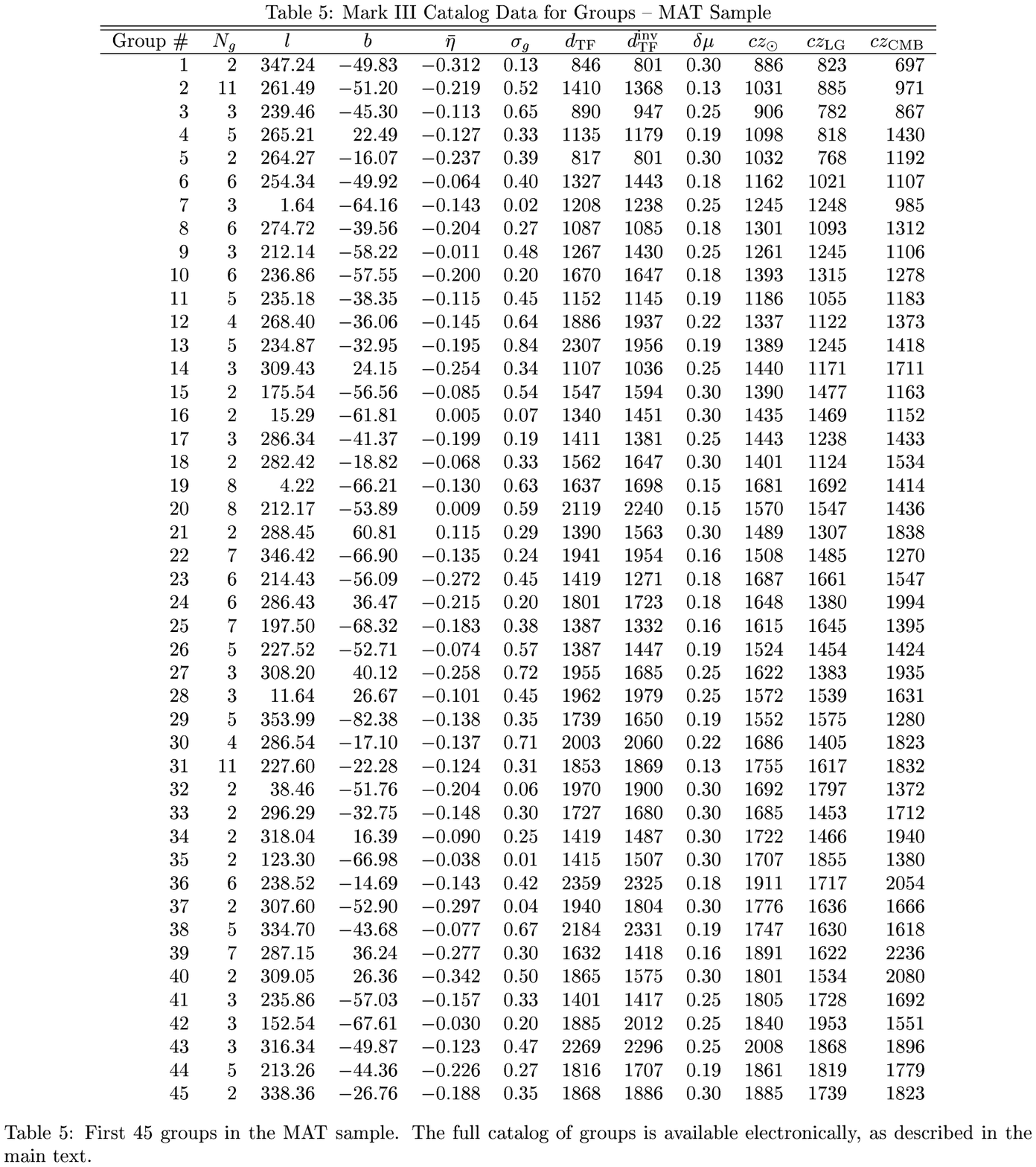.}}
\label{tab:matg}
\end{table}

\begin{table}[th]
\centerline{\begin{tabular}{l | c c r c r r r c l}
\multicolumn{10}{c}{Common-System Transformations for Mark III Samples} \\ \hline\hline
Sample&$a_0$&$a_1$&$a_2$&$\sigeta$&$N_\eta$&$b_0$&$b_1$&$\sigma_m$&$N_m$ \\ \hline
HMCL      &0.000&1.000& 0.000&  --  &   -- &      0.00& 0.000&   --  &  -- \\
W91CL     &0.000&1.000& 0.000& 0.013&   112&    $-1.31$& 0.000&  0.135& 184$^{(a)}$ \\
W91PP     &0.004&1.000& 0.000& 0.016&    74&    $-1.31$& 0.000&  0.135& 184$^{(a)}$ \\
CF$^{(b)}$&0.004&1.000& 0.000& 0.028&   135&    $-1.31$& 0.000&  0.135& 184$^{(a)}$ \\
          &0.011&1.069&$-0.663$& 0.027&   135&          &      &       & \\
MAT       &0.065&0.831& 0.000& 0.034&   113&    $-0.19$& 0.000&  0.130& 114 \\
A82       &0.016&1.000& 0.000& 0.039&   130&    $-0.92$&$-0.212$&  0.246& 130 \\ \hline
\end{tabular}}
\caption{Coefficients for transforming the Mark III magnitude
and velocity width data to a common system. The meaning of
the coefficients $a_0,$ $a_1$, $a_2,$ $b_0,$ $b_1,$ and $b_2$
is defined by equations~\ref{eq:etaequiv} and~\ref{eq:mequiv}.
The quantities $\sigeta$ and and $\sigma_m$ are the rms
dispersions resulting from, and $N_\eta$ and $N_m$ the number of
objects involved in, the least-squares fits used to determine
the transformation coefficients. Notes: $(a)$ W91CL, W91PP,
and CF were combined to obtain a common transformation for
the $r$-band magnitudes. $(b)$ Two width transformations,
one linear and one quadratic, are given for CF; see text
for further details.}
\label{tab:commonsys}
\end{table}

\begin{table}[th]
\caption{This table is contained in the file {\bf 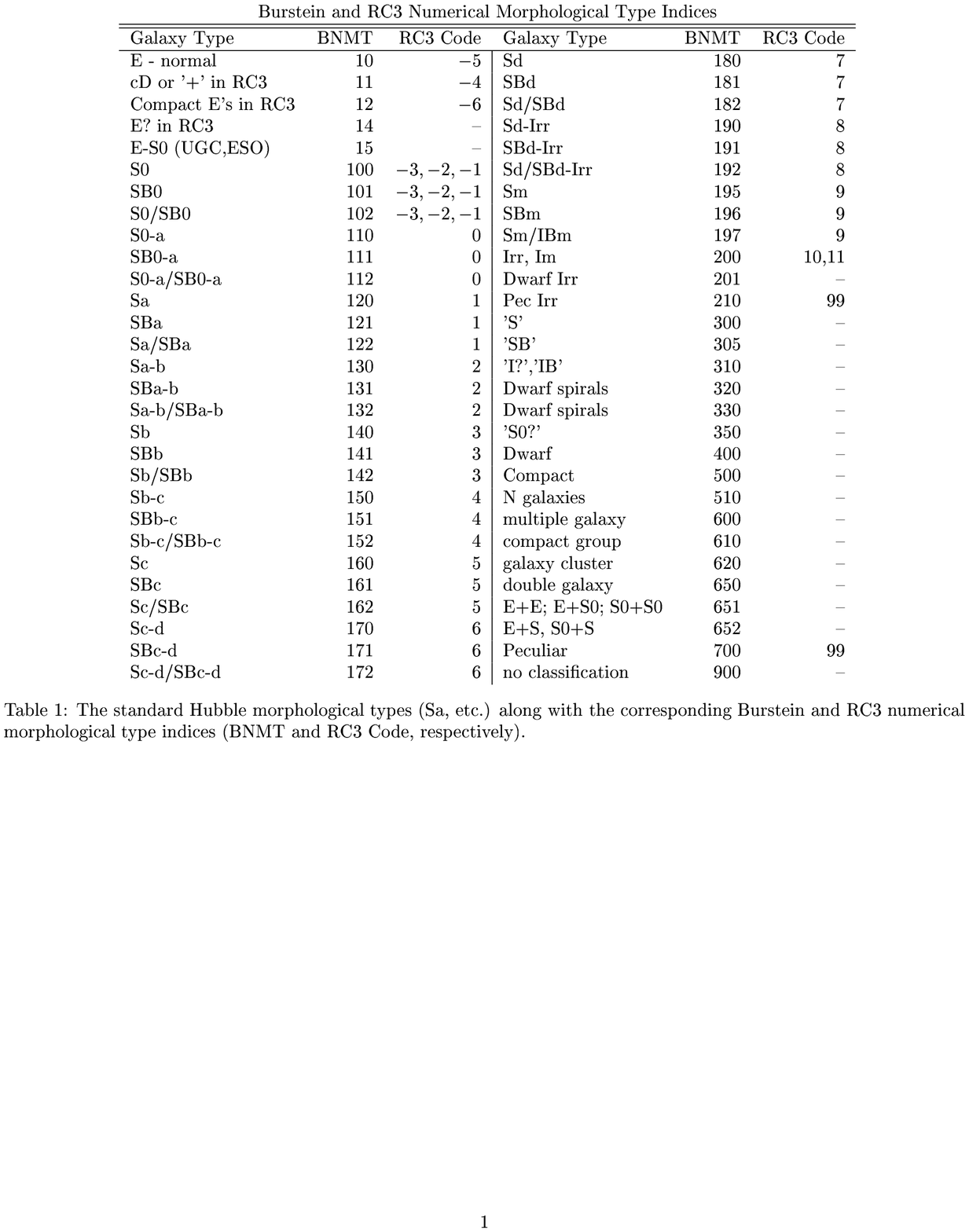.}}
\label{tab:morph}
\end{table} 

\end{document}